# Cluster Catalysis with Lattice Oxygen: Tracing Oxygen Transport from a Magnetite(001) Support onto Small Pt Clusters

Sebastian Kaiser,[1,2] Farahnaz Maleki,[3] Ke Zhang,[1,2] Wolfgang Harbich,[4] Ueli Heiz,[1,2] Sergio Tosoni,[3] Barbara A. J. Lechner,[1,*] Gianfranco Pacchioni,[3] Friedrich Esch[1,2]

[1] Chair of Physical Chemistry, Department of Chemistry, Technical University of Munich, Lichtenbergstr. 4, 85748 Garching, Germany

[2] Catalysis Research Center, Technical University of Munich, Lichtenbergstr. 4, 85748 Garching, Germany

[3] Dipartimento di Scienza dei Materiali, University of Milano-Bicocca, via Roberto Cozzi 55, 20125 Milano, Italy

[4] Institute of Physics, Ecole Polytechnique Fédérale de Lausanne, CH-1015 Lausanne, Switzerland

* bajlechner@tum.de

ABSTRACT. Oxidation catalysis on reducible oxide-supported small metal clusters often involves lattice oxygen. In the present work, we trace the path of lattice oxygen from $Fe_3O_4$(001) onto small Pt clusters during the CO oxidation, aiming at differentiating whether the reaction takes place at the cluster/support interface or on the cluster. While oxygen vacancies form on many other supports, magnetite maintains its surface stoichiometry upon reduction thanks to a high cation mobility. In order to investigate whether size-dependent oxygen affinities play a role, we study two specific cluster sizes, $Pt_5$ and $Pt_{19}$. By separating different reaction steps in our experiment, lattice oxygen can be accumulated on the clusters. Temperature programmed desorption (TPD) and sophisticated pulsed valve experiments indicate that the CO oxidation takes place on the Pt clusters rather than at the interface. Scanning tunneling microscopy (STM) shows a decrease in apparent height of the clusters, which density functional theory (DFT) explains as a restructuring following lattice oxygen reverse spillover.

1. **INTRODUCTION**

From fine chemical synthesis over combustion control to electrode design – about 80% of industrial chemical processes rely on catalysts to improve energy and material efficiency,[1] of which about 85–90% are catalyzed heterogeneously.[2] Such a heterogeneous catalyst typically consists of small metal nanoparticles (several 100s to 1000s of atoms) or clusters (often defined as ≤ 100 atoms) dispersed on oxidic support materials.[3,4] In contrast to the bulk, small clusters exhibit discrete electronic states, comparable to molecules, and therefore develop unique, strongly size-dependent chemical and physical properties. Small Pt clusters, for example, exhibit a non-metallic behavior.[5] These properties can even change the catalytic behavior: Au, which is inert in the bulk state, becomes highly active and selective for redox catalysis.[6–8] In small particles, a large proportion of atoms is in direct contact with the underlying support, which can strongly influence their catalytic performance and stability: It can alter the catalytic activity of the particles indirectly,[9,10] itself be an integral part of the catalytic cycle,[11] and even encapsulate the particle in the strong metal support interaction (SMSI),[12–14] thereby reducing[15] or enhancing[16] the catalytic activity.

In the case of the CO oxidation on supported metal clusters, several reaction mechanisms have been distinguished based on the oxygen supply: When clusters are supported on non-reducible oxides, the CO oxidation typically takes place in a Langmuir-Hinshelwood mechanism, where CO and $O_2$ co-adsorb from the gas phase and react.[17,18] On reducible oxide-supported clusters, on the other hand, the support can contribute lattice oxygen in a Mars van Krevelen (MvK) mechanism, thus not only providing a large oxygen reservoir but potentially also causing the entire catalyst system to restructure continuously.[19,20] The removal of lattice oxygen typically requires thermal activation, which can be facilitated in proximity to the clusters in a so-called metal-assisted MvK mechanism.[21] The resulting oxygen vacancies subsequently have to be refilled by gas phase $O_2$ in order to complete the catalytic cycle. A central question remains whether the oxidation step takes place at the cluster/support boundary or on the cluster. For larger nanoparticles on supports such as ceria[22,23] and $Co_3O_4$(111),[24] it was demonstrated that the reaction takes place on the particles via an oxygen reverse spillover, *i.e.* lattice oxygen transport onto the particles, which we will thus refer to as "lattice oxygen reverse spillover". We note that this process is different from the classical reverse spillover, where the partial pressure of the reactants may be increased through adsorption from the gas phase in a capture zone and migration to the catalytic site.[25]

Here, we study a similar CO oxidation catalyst with two important differences: We use small monodisperse clusters at the extreme low end of the size scale (namely $Pt_5$ and $Pt_{19}$) and we deposit them on a magnetite, $Fe_3O_4$(001), support, where we expect the oxygen exchange to be influenced by a high cation mobility: While oxygen vacancies created on the above-mentioned supports need to be refilled from the gas phase, magnetite can maintain its surface stoichiometry by facile iron diffusion into the bulk,[26] and thus provides a flexible oxygen reservoir. Magnetite is a magnetic, abundant, and reducible metal oxide which crystallizes in an inverse spinel structure with $Fe^{2+}$ occupying octahedral sites and $Fe^{3+}$ occupying tetrahedral and octahedral sites in a 1:1 ratio, within an fcc-$O^{2-}$ anion lattice.[27–29] Its (001) surface reconstructs into the subsurface cation vacancy (SCV) structure, whereby only $Fe^{3+}$ occurs in the uppermost layers, yielding a $(\sqrt{2} \times \sqrt{2})R45°$ reconstruction.[30] In the STM, this structure is observed in the form of parallel, undulating rows of iron atoms in octahedral sites, rotated 90° between adjacent terraces. The $Fe_3O_4$(001) surface is rich in surface defects such as surface hydroxyls, antiphase domain boundaries (APDB) between two reconstruction domains, or Fe-rich point defects such as unreconstructed unit cells and Fe adatoms.[31–34] These defects also act as adsorption and dissociation sites, thus participating in the surface chemistry.[35–37]

In the present work, we combine a multi-modal experimental approach comprising scanning tunneling microscopy (STM), temperature programmed desorption (TPD) as well as highly sensitive pulsed reactivity experiments with state-of-the-art density functional theory (DFT) calculations to investigate a $Pt_n/Fe_3O_4(001)$ model catalyst during exposure to a CO oxidation reaction environment. We demonstrate lattice oxygen reverse spillover onto the small Pt clusters, observe a size dependence of CO poisoning, and identify an oxygen-induced restructuring of the clusters.

## 2. METHODS

### 2.1. Experimental Methods

Naturally grown $Fe_3O_4(001)$ samples (SurfaceNet GmbH) were prepared by several preparation cycles, each consisting of 20 min $Ar^+$ ion sputtering at room temperature (4 x $10^{-5}$ mbar Ar, 1 keV, 3.6 µA sputter current), followed by annealing in 5 x $10^{-7}$ mbar $O_2$ at 983 K. This procedure yields a reproducible and clean surface, which was checked for stoichiometry as well as carbon impurities using X-ray photoelectron spectroscopy (XPS) and STM on a regular basis. For heating, a boron nitride heater located directly in the sample holder in contact with the sample was used; the temperature was measured by a type K thermocouple directly attached to the crystal.

Size-selected Pt clusters were generated by a laser ablation source.[38] Pt is evaporated from a rotating target using the second harmonic of a pulsed Nd:YAG laser; the resulting plasma is subsequently cooled in the adiabatic expansion of a He pulse (Westfalen AG, grade 6.0). This yields a supersonic beam of clusters, which is focused by several electrostatic lenses and guided towards a 90° quadrupole bender for positive charge selection. The resulting beam is mass-selected by a quadrupole mass filter and subsequently soft-landed on the substrate (i.e. with a kinetic energy < 1 eV/atom). The clusters were deposited at room temperature. For TPD and reactivity measurements, a cluster density of 0.05 clusters /nm$^2$ was used, for STM a lower cluster coverage of 0.01 clusters /nm$^2$ allowed optimal evaluation of the terrace background around each cluster. All experiments were carried out under ultrahigh vacuum (UHV) conditions, with a system base pressure of < 1 x 10$^{-10}$ mbar. All STM measurements were performed in constant current mode with a commercial Omicron VT-SPM instrument using electropolished Pt/Ir tips (Unisoku). STM image processing was performed with Gwyddion[39] using plane subtraction and row by row alignment tools for background correction. The height distribution of the particles was determined using a home-written Igor routine by detection of the particles via an intensity threshold, drawing a profile through the cluster maximum and determining the height of the cluster with respect to its surrounding background.

The experimental setup was recently equipped with a device for high sensitivity TPD and reactivity measurements, the so-called sniffer, which was built adapting a design by Bonanni *et al.*[40] In short, this device combines a pulsed reactant doser with a differentially pumped quadrupole mass spectrometer (Pfeiffer Vacuum GmbH, QMA 200 Prisma Plus). Up to two different reactants can be pulsed independently. The reactants as well as the desorption and reaction products are guided using a heated quartz tube assembly. As the distance between sniffer entrance and sample is typically in the range of only ~100 – 200 μm, the investigated surface is sufficiently decoupled from the rest of the chamber, thus a high amount of reactant can be dosed while keeping UHV conditions in the surrounding chamber. In the present study, each pulse corresponded to a dosage of approximately 0.1 L. This decoupling in combination with the guiding tubes results in the vast majority of the desorption and reaction products reaching the QMS, yielding a very high sensitivity, as well as the spectra being free of any additional peaks caused by desorption from e.g. the sample holder or the manipulator. The TPD and pulsed reactivity experiments were carried out using $O_2$ (Westfalen AG, grade 5.0) and $C^{18}O$ (Eurisotop, 96.1%). All reactivity and TPD-related samples have been saturated with $C^{18}O$ during cluster deposition. The delay between the pulses during the reactivity measurements was 4 s for $O_2$, 2.5 s for $C^{18}O$, and 2.5 s when pulsing $O_2$ and $C^{18}O$ alternatingly. The evaluation of the pulsed reactivity data was carried out using a Matlab-based evaluation tool that differentiates between pulse-related gas signals and background by detecting single peaks and evaluating in a pre-set time window. The thus-determined background comprises offsets, desorption signals, initial pulse valve transients and reactivity linked to long residence times and is subtracted; afterwards each pulse is integrated numerically. For details, see Figure S1 in the Supporting Information (SI).

### 2.2. Computational models and methods

DFT calculations have been performed to model the lattice oxygen reverse spillover process. To this end, we considered a $Pt_5$ cluster supported on the $Fe_3O_4$(001) surface. For the calculations we used the code VASP 5.[41] The core electrons are described with the projector-augmented wave method.[42,43] O(2s, 2p), Fe(3s, 3p, 3d, 4s) and Pt(5d, 6s) electrons are treated explicitly. Spin-polarized calculations are carried out at the level of the Generalized Gradient Approximation (GGA) adopting the Perdew, Burke and Ernzerhof PBE exchange-correlation functional.[44] The strongly correlated character of Fe(3d) electrons is accounted for as in the GGA+U approach[45,46] by applying a Hubbard U parameter of 3.8 eV to the Fe 3d states.[30] The magnetic structure of $Fe_3O_4$ is correctly reproduced. Long-range dispersion is included according to the D3 method introduced by Grimme.[47] The relaxation of the magnetite bulk lattice parameters and internal coordinates has been performed with a 3x3x3 K-points grid and a kinetic energy cutoff of 600 eV. All subsequent calculations on the (001) surface have then been carried out with a 2x2x1 K-points grid and a kinetic energy cutoff of 400 eV. The calculations on gas phase Pt clusters are carried out in Γ point only in a cubic 25 Å box. The truncation criteria for the electronic and ionic loops are $10^{-5}$ eV and $10^{-2}$ eV/Å, respectively. Kinetic barriers are calculated according to the Climbing Image-Nudged Elastic Band approach,[48] considering four images along the reaction path.

The (001) surface of the magnetite is modelled by a ($\sqrt{2}$ x $\sqrt{2}$)R45° $Fe_{35}O_{48}$ slab, according to the subsurface-vacancy model formerly proposed by Parkinson and co-workers.[30] Further details on the model structures are reported in Section S2 of the SI.

To model the formation of an O vacancy, we distinguish three non-equivalent oxygen lattice positions on the $Fe_3O_4$(001) surface. O1 and O2 lie on threefold-coordinated surface sites connecting Fe atoms in tetrahedral positions. O3 lies on top of an Fe ion in an octahedral site (see Figure S4). The thermodynamic stability of the O vacancy (estimated with respect to the pristine surface and ½ $O_2$) depends on the site where the oxygen atom is removed. In particular, O3 located on top of Fe ions in octahedral sites can be removed at a cost of 2.91 eV; removal of O2 (3.10 eV) and O3 (3.55 eV) requires more energy (see also SI, Section S3).

## 3. RESULTS AND DISCUSSION

In order to identify how CO binds to $Pt_n$/$Fe_3O_4$(001) and whether it reacts with lattice oxygen, we start with a discussion of the TPD spectra of CO-saturated surfaces without external oxygen supply. Subsequently, the reactivity linked with either the CO or $O_2$ reactant is monitored by pulsed reactivity measurements. A low temperature reaction peak demonstrates that the lattice oxygen reverse spillover leads to a highly reactive oxygen species on the cluster. Finally, we will show that the lattice oxygen reverse spillover leads to structural changes of the clusters. We should note at this point that over the entire temperature range discussed below, STM indicates the cluster coverage remains constant and the distribution monodisperse. In the following, ripening can thus be excluded from the discussion.

**3.1. Setting the stage: CO oxidation in TPD measurements**

Figure 1(a) shows CO desorption curves from $Pt_5$ (orange) and $Pt_{19}$ clusters (gray) deposited onto the $Fe_3O_4$(001) surface. First, $Pt_{19}$ clusters exhibit a broad desorption feature with an onset temperature of around 350 K and a peak temperature of 475 K. This feature is not observed on the bare magnetite surface (see Figure S5 in the SI), agrees with the temperature range reported for CO desorption from Pt nanoparticles on $Fe_3O_4$(001),[49] and can therefore be assigned to the clusters. Second, the weak and broad feature with a maximum at 710 K is similar to the CO recombinative desorption from Fe(111).[50] We thus conclude that it results from CO which has been adsorbed dissociatively on the small number of highly reduced iron sites, most likely Fe adatoms, which are a common defect on the $Fe_3O_4$(001) surface.[31] Finally, there is another small feature around 330 K. Such features have been tentatively assigned in literature to reduced iron sites that surround Pt nanoparticles on $Fe_3O_4$(111), when strong metal support interaction sets in.[51] While this interaction requires higher temperatures not yet reached here, there could already be clusters present in close proximity to reduced iron defects.

The CO desorption features of $Pt_5$ are comparable to those of $Pt_{19}$, with the difference that the desorption maxima are shifted to higher temperatures, indicating stronger binding. Note that all $Pt_5$ measurements end at 625 K, as cluster ripening was observed in STM when heating to higher temperatures (see Figure S6). The main CO desorption feature from $Pt_5$ with an onset around 420 K and a peak at 550 K is assigned to the clusters. A lower temperature feature at around 430 K is observed, comparable to the one observed at 330 K for $Pt_{19}$. For both cluster sizes, this feature is located around the onset temperature of the main desorption peak, i.e. both features are shifted similarly. This might indicate a similar origin of the lower temperature feature, namely due to clusters in the vicinity of reduced surface defect sites.

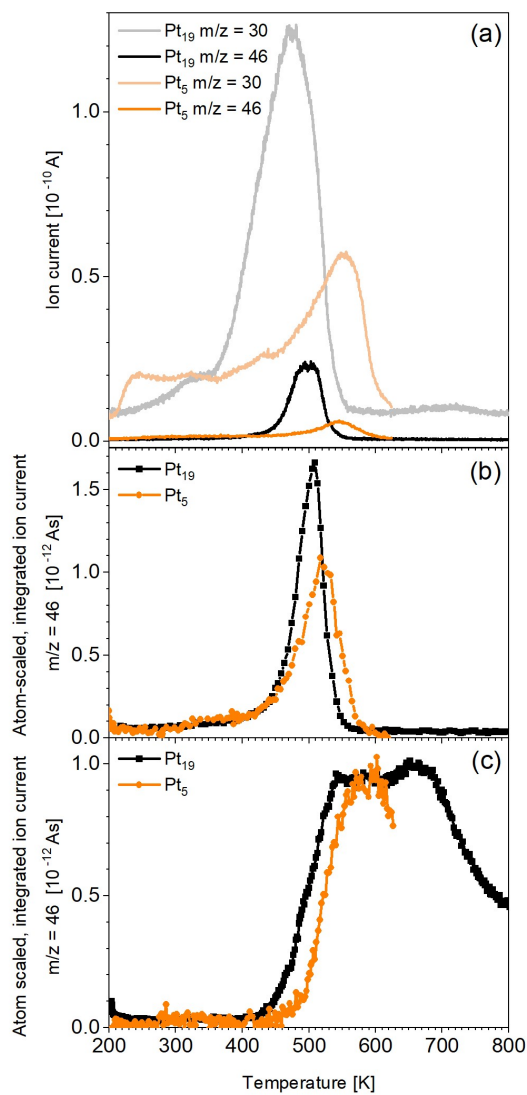

**Figure 1.** TPD and pulsed reactivity spectra of $Pt_{19}$ (black, grey) and $Pt_5$ (orange) supported on $Fe_3O_4(001)$. By using the same cluster density of 0.05 clusters/nm$^2$, the resulting atom density is ~4 times higher for the $Pt_{19}$ sample. (a) Saturation TPD curves of $C^{18}O$ adsorbed at 200 K. The $C^{18}O$ (m/z = 30) and $C^{18}O^{16}O$ (m/z = 46) traces are shown. (b) $C^{18}O^{16}O$ (m/z = 46) production as a function of the temperature, synchronized with $O_2$ reactant pulses on $C^{18}O$-saturated samples. (c) $C^{18}O^{16}O$ (m/z = 46) formation synchronized with $C^{18}O$ reactant pulses on $C^{18}O$-saturated samples. The heating rate was 1 K/s. All pulsed data were integrated and normalized to the Pt atom coverage.

In both samples, we observe weak desorption signals in the temperature range < 300 K, which we assign to desorption from defects in the magnetite support (see TPD for bare magnetite in Figure S5).[35] Since their intensity decreases upon cluster deposition, we conclude that a significant part of the clusters is most likely located on surface defect sites such as unreconstructed unit cells or antiphase domain boundaries.[31,52] Note that the background at these low temperatures varies with the state of the single crystal which we suspect is due to slight changes in sample cleanliness, oxidation state and/or defect density. The measurements comparing bare and cluster-covered magnetite were therefore recorded in close temporal sequence.

Having understood the CO features, we can now interpret the $C^{18}O^{16}O$ production signals recorded simultaneously with the CO desorption (also shown in Figure 1(a)). Since no oxygen is dosed, this $CO_2$ production, observed for both cluster sizes, must be strictly related to the availability of lattice oxygen for reaction with CO. During the TPD measurements, only $C^{18}O^{16}O$ and no $C^{18}O_2$ is observed, excluding a Boudouard-type reaction (2 CO → C + $CO_2$). We are thus observing a MvK reaction. For $Pt_{19}$, the $CO_2$ feature has an onset temperature located around 400 K and a broad peak at 500 K. $Pt_5$ exhibits a similar feature, starting below 420 K with the peak at 545 K. Note that we observe a change in apparent cluster height in the same temperature range in STM measurements, as will be discussed later. As no $CO_2$ desorption could be detected for the bare $Fe_3O_4$(001) surface (see Figure S5), we attribute the $CO_2$ signals to the presence of the clusters. For both cluster sizes, they exhibit about the same onset temperature and peak position with respect to the corresponding CO desorption peaks, indicating that the $CO_2$ formation is closely related to the CO desorption.

## 3.2. Exploring the reaction mechanism by overcoming reactant limitations: Pulsed-reactivity measurements

Complementary to the TPD experiments, we performed pulsed valve experiments that give access to the reaction rate synchronized with a given reactant pulse. In these background-corrected measurements (see Section S1), each data point is the integral over a product pulse and correlates with the reaction rate at a given temperature. These reactant-synchronized pulsed measurements allow us to investigate adsorption limitations, (i) of oxygen adsorption by CO poisoning (in pulsed $O_2$ experiments) and (ii) of CO adsorption at high temperatures (in pulsed CO experiments).

We start with pulsing $O_2$ onto the CO pre-covered sample and observe the $CO_2$ production shown in Figure 1(b) for $Pt_5$ (orange) and $Pt_{19}$ clusters (black). The $CO_2$ formation starts around 400 K for both cluster sizes and peaks at 505 K for $Pt_{19}$ and 520 K for $Pt_5$, respectively. This difference reflects the slightly higher binding energy of CO on $Pt_5$. While the cluster surface is poisoned with CO at low temperatures, binding sites for dissociative $O_2$ adsorption become available above the onset temperature of CO desorption, leading to co-adsorption of oxygen and CO on the same cluster.[18] At this temperature, the CO oxidation in a classical Langmuir-Hinshelwood fashion is already facile on clusters[53] and hence a high reaction rate is observed as soon as the adsorption limitations are overcome. The similarity to the $CO_2$ production in the TPD indicates that it is adsorption-limited as well. As we will discuss in detail in Section 3.5, DFT confirms that the activation barrier for the initial lattice oxygen reverse spillover is below 1 eV, hence far less than the CO binding energy calculated e.g. for $Pt_2$ clusters on $Fe_3O_4(001)$.[54]

We note that the $CO_2$ production peak on $Pt_5$ in the pulsed experiments is at a somewhat lower temperature than the peak observed in TPD, contrary to $Pt_{19}$. This observation can be explained by a reaction limitation due to a lack of CO which is not replenished during the measurement. When supplying CO by alternating CO and $O_2$ pulses (see Figure S7), the $CO_2$ formation synchronized with the $O_2$ pulses follows the shape of the corresponding $CO_2$ TPD trace, with the same onset and peak temperatures, thus confirming the hypothesis above.

We now describe the opposite experiment, pulsing CO without providing any oxygen. Figure 1(c) shows the $CO_2$ production observed for both investigated cluster sizes as a function of temperature. For $Pt_{19}$, $CO_2$ formation starts at around 425 K, reaching a plateau between 550 K and 670 K with approximately constant $CO_2$ formation rate. On $Pt_5$, an onset temperature of 450 K is observed, with a maximum at 590 K. This experiment allows to probe the behavior of the $CO_2$ formation over a wider temperature range, since the reactant is refreshed continuously by pulses and thus also available beyond the CO desorption temperature. As expected, the $CO_2$ production starts at the onset temperature of $CO_2$ formation in the TPD experiment, for both cluster sizes, although at a slightly higher temperature, where oxygen migration onto the cluster is no longer hindered by CO. The signal saturates once all available cluster sites can accommodate active oxygen atoms. This temperature coincides with the temperature where CO desorption is complete. Interestingly, the normalized peak intensity is the same for both cluster sizes. The number of active sites hence scales with the surface and not the rim of the cluster.[1] The upper limit of the plateau could be due

---

[1] The intensity shown in Figure 1 (c) is normalized to the number of Pt atoms per cluster. Taking the average cluster heights at ≥ 450 K from our STM measurements (see Section 3.4) into account for a conservative geometrical estimate, we can estimate the number of atoms at the rim and on the surface of the clusters and normalize them to the number of cluster atoms: $Pt_5$ is a single layer cluster with 5 (normalized 1) rim atoms and 5 (normalized 1) surface atoms (see Figure 4(c)) and $Pt_{19}$ a bilayer cluster with ~9 (normalized 0.5) rim and ~16 (normalized 0.8) surface atoms,

to encapsulation effects or due to limited oxygen coverage: chemisorption on stepped Pt(112) leads to $O_2$ desorption just below 700 K.[55] All the plateau characteristics point to a reaction that is limited by free adsorption sites on the cluster surface and that proceeds *on* the cluster, via lattice oxygen reverse spillover. In contrast, an interface reaction would not saturate as long as CO (and lattice oxygen) are available since the two reactants do not compete for adsorption sites.

### 3.3. Lattice oxygen reverse spillover in sequential CO TPDs

Having shown the ease of lattice oxygen reverse spillover once adsorption sites are available, we now use this highly active oxygen species in a low temperature reaction. We know from literature that the CO oxidation on Pt clusters after sequential adsorption of oxygen and CO already takes place below 400 K.[18,53]

In the experiment in Figure 2, we show that it is indeed possible to observe an additional $CO_2$ production feature when populating the clusters with oxygen by reverse spillover prior to CO adsorption. Figure 2 shows a sequence of $CO_2$ production TPD curves on a $Pt_5$ sample, each taken after saturating the surface with CO at 200 K. Remarkably, the anticipated low-temperature $CO_2$ feature around 300-370 K is observed only for one out of four runs, the second one. While CO poisoning and an activation barrier hinder oxygen migration onto the cluster in the first run, the second run starts already with a certain oxygen coverage on the clusters, built up from lattice oxygen during the first temperature ramp. This reaction is, however, not repeated in the third and fourth run, where another effect comes into play that is clearly reflected in the intensity loss of the primary $CO_2$ production peak: At the high temperature end of the TPD, cluster encapsulation due

---

respectively. The number of active sites hence scales with the surface and not the rim of the cluster as we see similar activities per cluster atom.

to SMSI starts, an effect that strongly affects the cluster reactivity, as studied for nanoparticles,[49] while requiring further investigation for clusters. Here, it leads to the well-known loss of CO adsorption capability that we see in an independent series of CO TPDs (see Figure S8). Since the encapsulation is incomplete after the first run, but finished after the second, we no longer observe the facilitated low-temperature $CO_2$ production thereafter. Note that the low intensity of this $CO_2$ production feature is likely caused by partial encapsulation after the first ramp. Summing up, this experiment provides direct proof for reverse spillover of lattice oxygen.

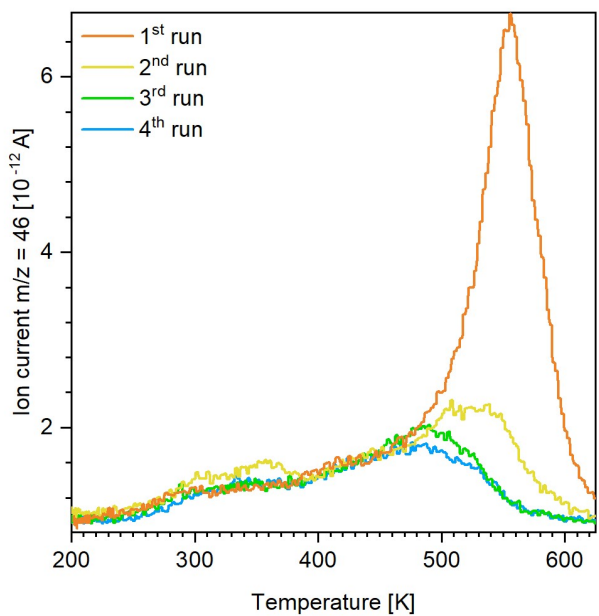

**Figure 2.** $C^{18}O^{16}O$ (m/z = 46) formation in subsequent CO-TPD spectra of $Pt_5$ supported on $Fe_3O_4$(001). In each run, the surface is saturated with $C^{18}O$ at 200 K; the heating rate was 1 K/s. Only in the second run (yellow), a $CO_2$ production feature around 300-370 K is observed. The main feature above 500 K decreases during the first two runs. Figure S8 shows the corresponding $C^{18}O$ desorption signals.

## 3.4. Effects of the lattice oxygen reverse spillover on cluster structure: STM measurements

In the following, we focus on the effects of this lattice oxygen reverse spillover on the cluster structure, concerning geometry and electronic state. Figure 3(a–e) shows a series of STM images displaying $Pt_{19}$ clusters on a defect-rich $Fe_3O_4(001)$ surface, recorded at the temperatures indicated in the Figure. The clusters appear as bright protrusions on the surface and seem to be largely randomly distributed while maintaining their size upon deposition. At all temperatures investigated here, the clusters are still monodisperse; ripening or disintegration is not observed. At 473 K and even more pronounced at 573 K, tiny holes form around some of the clusters, but not all (details see Figure S9). Such holes are typically observed on this surface when oxygen atoms are removed, leaving behind undercoordinated iron atoms that diffuse into the bulk.[20] This is a direct consequence of the lattice oxygen reverse spillover occurring around the clusters. The hole formation is more or less pronounced depending on the amount of CO in the chamber background that reacts off oxygen from the clusters, creating free adsorption sites for renewed oxygen migration.

Upon annealing, the cluster brightness seems to decrease, indicating a decline in their apparent height, while the coverage remains constant (confirmed by statistical analysis of several images of the same sample at different temperatures). This finding is quantified in Figure 3(f) for $Pt_{19}$ where the cluster height distributions at different temperatures are presented in histograms. Dashed red lines are used to indicate the approximate correspondence to atomic layers. While the $Pt_{19}$ clusters are about three atomic layers high at room temperature, they flatten to between one and two layers at 573 K. The transition takes place between 423 K and 523 K, which is the temperature range where $CO_2$ formation is observed in Figure 1(a). This strong correlation in temperature suggests that the change in apparent height is caused by lattice oxygen reverse spillover.

When comparing several cluster sizes in Figure 3(g), we find that this is a general phenomenon. The decrease in height holds for all investigated cluster sizes, suggesting the lattice oxygen reverse spillover to be a universal effect for the $Pt_n/Fe_3O_4(001)$ system. Such a behavior could either be explained by a real geometry change of the clusters or by a change in the density of states due to oxygen adsorption.

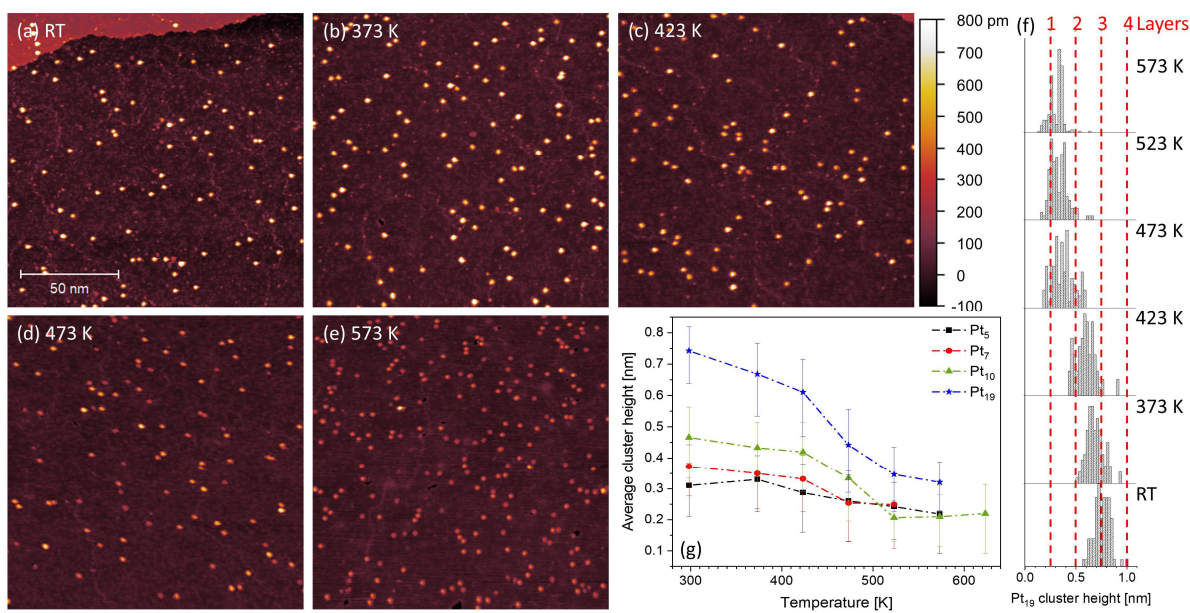

**Figure 3.** Height evolution of $Pt_n/Fe_3O_4(001)$ as a function of temperature. (a)–(e) STM images (150 x 150 nm$^2$) of $Pt_{19}$ clusters (0.01 clusters/nm$^2$) measured at the temperatures indicated. All images have been scaled to the same false color scale for height comparison.[2] (f) Height profiles comparing the size distribution of $Pt_{19}$ in (a)–(e) as a function of temperature. (g) Evolution of the average cluster height as a function of temperature for $Pt_5$, $Pt_7$, $Pt_{10}$ and $Pt_{19}$. *Imaging parameters:* $V_b = 1.50$ V; (a, e) $I_t = 300$ pA, (b, c, d) $I_t = 400$ pA.

---

[2] Note that (e) is from a second set of experiments with a slightly higher cluster coverage than (a)-(d). The coverage in (e) is consistent with the coverage of the same sample after RT deposition.

**3.5. DFT calculations reveal further mechanistic details**

The experimental results point to some mechanistic aspects that require theoretical investigation. In particular, we consider the following questions: (i) How easy is it to remove lattice oxygen from bare magnetite vs. in the vicinity of a cluster? (ii) Is lattice oxygen reverse spillover endo- or exothermic and can we confirm that the clusters can be covered by considerable amounts of oxygen? (iii) What is the activation barrier? (iv) Can the apparent height change in STM be explained by a restructuring?

First, we studied the structure of a gas phase $Pt_5$ cluster, starting from the most stable isomers reported in the literature.[56–62] This led to six isomers with very similar stability; the energy difference between the ground state and the least stable of the six isomers considered here is 0.35 eV, and sometimes different structures are separated by a few meV, suggesting a fluxional behavior of the gas phase clusters (see SI, Section S9). Upon deposition on $Fe_3O_4$, some of the structures retain the topology of the gas phase, while others undergo a strong rearrangement. The $Pt_5$ adsorption energies (Table S4), computed with respect to $Fe_3O_4(001)$ and the most stable $Pt_5$ isomer ($Pt_5(i)$ in Figure S10), show that the ground state corresponds to a capped rhombus, with an adsorption energy of -4.51 eV, shown in Figure 4(a), followed by a square pyramidal $Pt_5$ cluster (Figure 4(b)) and a planar structure (Figure 4(c)). Note that the most stable isomer, Figure 4(a), is 1.5 eV more favorable than the next one, which shows that the surface has a strong stabilizing effect on this structure. This also suggests that the conversion of one isomer into another is much more difficult than in the gas phase.

Upon deposition, the Pt cluster becomes slightly positively charged (by 0.3 e) and a partial quenching of the magnetic moment of the support occurs, with a reduction from 3.3 $\mu_B$ per $Fe_3O_4$ formula unit in the clean surface, to 2.59 $\mu_B$ when $Pt_5$ is adsorbed, as shown in Table S4.

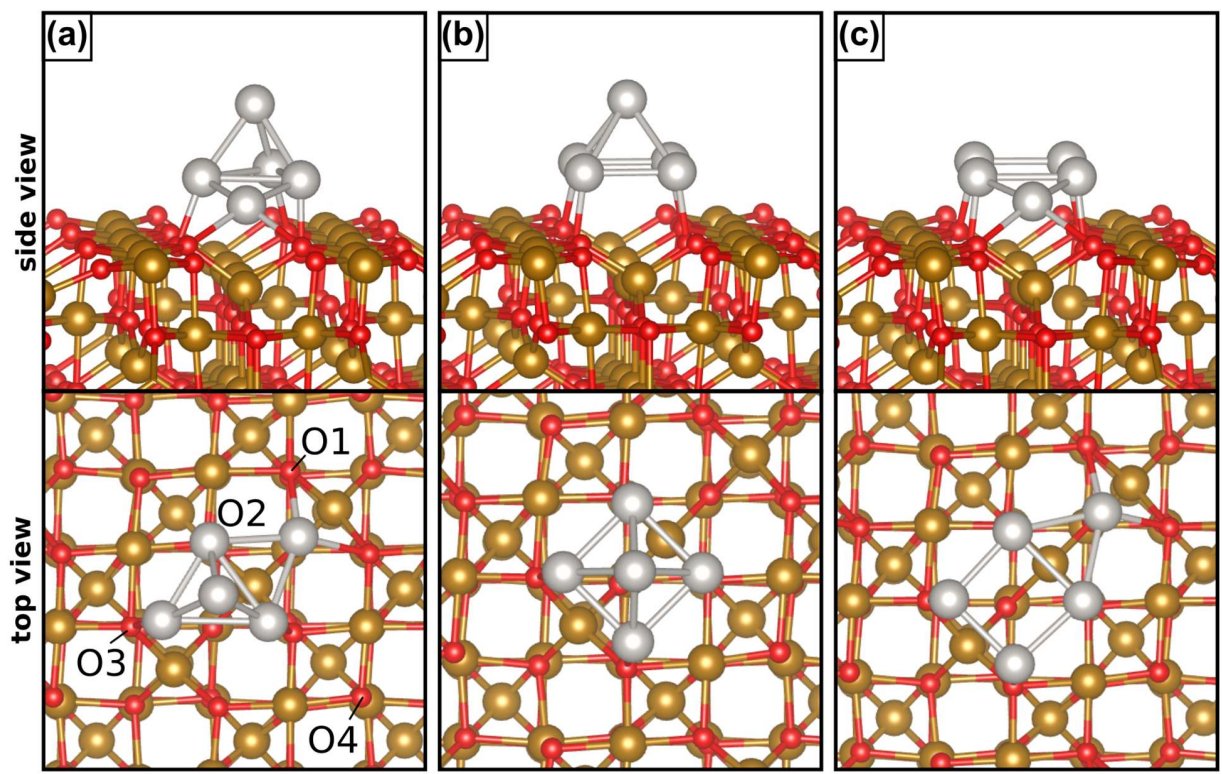

**Figure 4.** $Pt_5$ clusters on the $Fe_3O_4(001)$ surface. (a), (b) and (c) are the three most stable isomers, with isomer (a) being 1.5 eV and 1.78 eV lower in energy than the isomers in (b) and (c), respectively. Red, brown, and grey spheres correspond to O, Fe and Pt atoms, respectively.

We now start addressing question (i). The formation energy of an O vacancy on the clean $Fe_3O_4(001)$ surface is about 3 eV (see Table S2 and Section S3). The formation of a metal/oxide interface may facilitate the removal of the O atoms in contact to the metal adduct.[21] We have explored this possibility by removing O atoms from four possible sites, indicated in Figure 4(a), involving O either in direct contact with $Pt_5$, or at some distance. The formation energy of an O vacancy on $Pt_5/Fe_3O_4$ is reduced with respect to the clean surface. In particular, the formation of $V_{O2}$ has a cost of 1.98 eV, about 0.9 eV smaller than the most favorable case on the clean surface, which costs 2.91 eV. Also, the removal of an O atom not in direct contact with $Pt_5$ (such as $V_{O4}$) is slightly easier by almost 0.5 eV (see also SI, Section S10). These results point to a metal-assisted MvK mechanism.

In a second step, we address question (ii) regarding the endo-/exothermicity of the reverse spillover. The mechanism of lattice oxygen reverse spillover involves the displacement of O atoms from the lattice positions and their adsorption on the supported metal cluster. To model this process, we have started from the most favorable case of O removal ($V_{O2}$ in Figure 4(a)) and have re-adsorbed an O atom on various sites of $Pt_5$. A preliminary exploration of O adsorption on gas phase $Pt_5$ provides information about the most stable adsorption sites (see Table S3). O adsorbs preferentially in a bridge mode or as a terminal Pt-O group, followed by threefold hollow sites. When the cluster is adsorbed on $Fe_3O_4$, other adsorption sites are present at the $Pt_5/Fe_3O_4$ interface. In some cases, the lattice oxygen reverse spillover process is endothermic, but there are also structures where the product is more stable than the initial configuration. The most favorable case is shown in Figure 5(a), where an O from the interface is inserted into a Pt-Pt bond, which is preferred by -0.40 eV (see Table S6 and Figure S11 for further details). Thus, lattice oxygen reverse spillover involving a single O atom is thermodynamically favorable.

This leads us to question (iii) where we want to connect the lattice oxygen vacancy formation with the oxygen adsorption on the cluster by a viable path. This process is activated and the kinetic barrier has been estimated for the initial step of the reaction, i.e. the displacement of an O atom from an O2 lattice site (see Figure 4(a)) to a nearby interfacial site where it binds at the cluster/magnetite interface, shown in Figure 5. For this specific path, the whole process is almost thermoneutral, and the barrier of 0.93 eV can be overcome by thermal effects at the temperatures involved in the experiments described above. It is reasonable to assume that the displacement of the O atom from the lattice is the rate-determining step of the whole process: further diffusion of

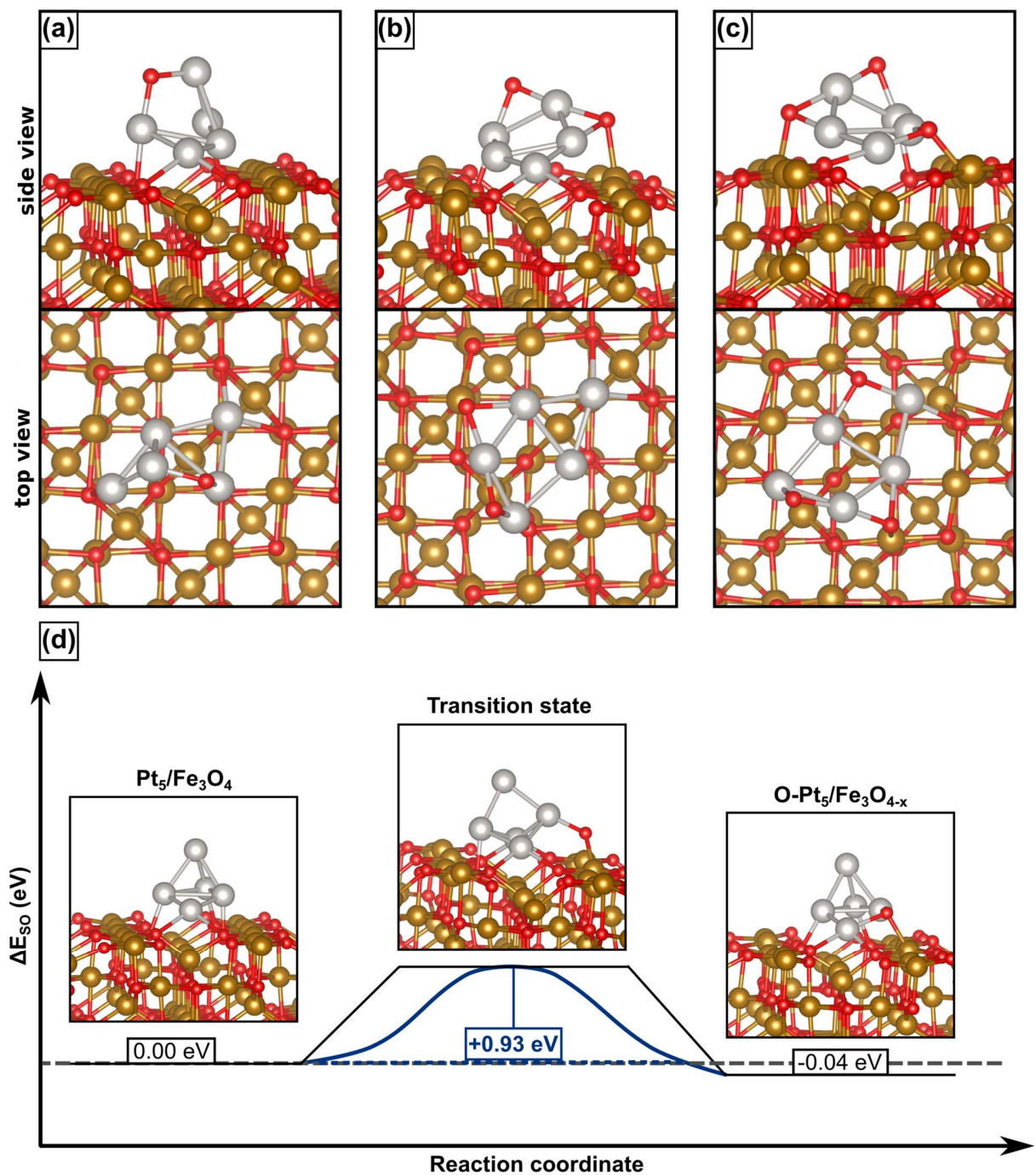

**Figure 5.** Side and top views of the most stable structures obtained for reverse spillover of (a) one, (b) two, and (c) three lattice oxygen atoms. In (d) the energy profile to displace an O atom from a lattice position to Pt$_5$ is shown.

O-species on $Pt_5$ should occur with low barriers due to similar bond strengths of various adsorption sites.

Next, we have considered the case of the removal of a second O atom from various lattice sites and its re-adsorption on $Pt_5O$. Also in this case, we found a few structures where the process is exothermic. The formation of the most stable isomer, shown in Figure 5(b), where two O atoms are adsorbed on a bridge site, is exothermic by -0.64 eV (see also Table S7 and Figure S12). Finally, the process has also been considered for a third O atom which is removed from the cluster/oxide interface and re-adsorbed on $Pt_5$, forming an oxidized $Pt_5O_3$ supported cluster, as shown in Figure 5(c). This generates a large number of possible initial combinations, and we could explore only a few of them. In the most favorable case (Figure 5(c)), the transfer of three O atoms is slightly endothermic, by 0.17 eV only. We cannot exclude the existence of even more favorable isomers, but this already supports the evidence that multiple lattice oxygen reverse spillover from magnetite onto a Pt cluster is energetically accessible (for further details see Figures S11, S12 and Tables S7, S8).

This opens the question of the best channel for CO oxidation, via CO reaction with an O atom of $Fe_3O_4$ at the rim of $Pt_5$ or with the O atoms adsorbed on the Pt cluster. The experiments show a complex behavior as a function of CO poisoning, reaction temperature, cluster size, etc. With our simplified models, we compared the energy required to remove oxygen from the $Pt_5/Fe_3O_4$ interface or from the oxidized $Pt_5$ cluster. For $Pt_5/Fe_3O_4$, the removal of one O atom in contact with the cluster costs 2.0, 2.5, 2.7 and 3.1 eV, depending on the site (on average 2.6 eV). These numbers can be compared to the cost of desorbing an O atom from $Pt_5O$ supported on $Fe_3O_{4-x}$ where an O vacancy has been created. In the most stable configuration, this cost is 2.4 eV, indicating that the Pt-O bond is strong and that at low O coverage, removing the O ad-atoms from a Pt cluster has a comparable cost to removing O atoms at the cluster/oxide interface. Things change for higher O coverages though. We have removed each of the three O atoms from the $Pt_5O_3/Fe_3O_{4-3x}$ system, and the cost is 1.7, 2.2, and 2.2 eV, respectively (on average 2.05 eV). This suggests that for more pronounced lattice oxygen reverse spillover, or higher O coverages on Pt clusters, the reaction of CO with adsorbed oxygen becomes preferred.

Finally, in question (iv), we want to rationalize the structural changes observed in the STM images. Lattice oxygen reverse spillover indeed induces a structural change in the supported cluster, as apparent in Figures 4 and 5. In particular, there is a tendency to rearrange the structure of $Pt_5$ from 3D to a quasi-2D as the O loading increases. To quantify this effect, we have computed the average cluster height with respect to the $Fe_3O_4$ surface for all the cases discussed here, and indeed there is an overall reduction of the cluster height (see SI, Section S12). This could indicate that the height evolution observed in STM, as shown in Figure 3, is due to geometry changes and not purely an electronic effect. However, more work is necessary to confirm this conclusion.

Furthermore, Fe spillover from $Fe_3O_4$ onto $Pt_5$ and $Pt_5O_3$ has been considered, following a similar strategy as described above for lattice oxygen reverse spillover. In all cases considered, however, Fe reverse spillover is highly unfavorable (see SI, Section S13). Thus, if Fe diffusion occurs, this will be towards the bulk of magnetite, not towards the supported Pt clusters. These results suggest that the encapsulation of Pt clusters due to SMSI effects follows more complex paths than the diffusion of isolated Fe atoms.

## 4. CONCLUSIONS

A central question concerning cluster catalysis with lattice oxygen, in a Mars van Krevelen mechanism, is whether the reaction occurs at the cluster-support rim or on the cluster. We studied this distinction on the example of metal-assisted MvK in the CO oxidation on small, size-selected $Pt_n$ clusters supported on magnetite $Fe_3O_4(001)$ in a multi-modal experiment, supported by DFT calculations. The detailed TPD and pulsed reactivity experiments revealed a reaction on the cluster that involves reverse spillover of lattice oxygen, a phenomenon hitherto unknown on this support. Via annealing-induced migration, lattice oxygen can be accumulated on the clusters and react in a low temperature window of 300–370 K.

As our calculations for $Pt_5$ clusters showed, this reverse spillover process is exothermic for the first two oxygen atoms, with an initial migration activation barrier below 1 eV, much lower than the CO desorption barrier. Thus, the lattice oxygen reverse spillover is driven by overcoming CO poisoning and scales with the availability of free adsorption sites. The maximum obtainable turnover rates observed in the pulsed reactivity experiments scale with the number of cluster surface atoms.

STM investigations showed that the clusters remain monodisperse throughout all experiments, but with a distinct decrease in apparent height, concomitant to the formation of holes due to lattice oxygen removal. The calculations suggest that this change could be due to a true geometrical adaptation upon lattice oxygen reverse spillover resulting in a transition from 3D to 2D clusters.

Larger Pt nanoparticles on the same support have been shown to become encapsulated by a thin, reduced iron oxide film due to SMSI.[49] In first experiments we found this effect to also occur on small clusters, at temperatures where lattice oxygen reverse spillover already occurred. Calculations suggest that the growth of thin oxide films occurs through more complex paths than the diffusion of isolated Fe atoms, warranting more extensive future investigations.

**Supporting Information.** S1. Pulsed reactivity data evaluation. S2. Characterization of magnetite bulk and (001) surface by DFT calculations. S3. Formation of a surface oxygen vacancy on various sites. S4. CO desorption from bare magnetite and magnetite-supported clusters in TPD. S5. Ripening of $Pt_5$ in STM. S6. Alternating CO and $O_2$ pulses on $Pt_5/Fe_3O_4(001)$. S7. Loss of CO adsorption capability upon cluster encapsulation. S8. Hole formation in cluster vicinity. S9. Platinum clusters in the gas phase: structure relaxation and oxygen adsorption. S10. Adsorption of $Pt_5$ on $Fe_3O_4(001)$ and formation of oxygen vacancies. S11. Lattice oxygen reverse spillover. S12. Impact of lattice oxygen reverse spillover on cluster height. S13. Iron spillover.

## ACKNOWLEDGEMENTS

The experimental work was funded by the Deutsche Forschungsgemeinschaft (DFG, German Research Foundation) under Germany´s Excellence Strategy – EXC 2089/1 – 390776260 and project numbers ES 349/5-2 and HE 3454/23-2. This project has received funding from the European Research Council (ERC) under the European Union's Horizon 2020 research and innovation programme (grant agreement No 850764). B.A.J.L. gratefully acknowledges financial support from the Young Academy of the Bavarian Academy of Sciences and Humanities. F.M., S.T. and G.P. acknowledge support from the Italian Ministry of University and Research (MIUR) through the PRIN Project 20179337R7, the grant Dipartimenti di Eccellenza - 2017 "Materials For Energy", and the CINECA supercomputing center via ISCRAB.

## ORCID

Barbara A. J. Lechner: 0000-0001-9974-1738
Ueli Heiz: 0000-0002-9403-1486
Gianfranco Pacchioni: 0000-0002-4749-0751
Friedrich Esch: 0000-0001-7793-3341

*Sci.* **2003**, *525* (1–3), 66–84.

Supporting Information for

# Cluster Catalysis with Lattice Oxygen: Tracing Oxygen Transport from a Magnetite(001) Support onto Small Pt Clusters

*Sebastian Kaiser,[1,2] Farahnaz Maleki,[3] Ke Zhang,[1,2] Wolfgang Harbich,[4] Ueli Heiz,[1,2] Sergio Tosoni,[3] Barbara A. J. Lechner,[1,*] Gianfranco Pacchioni,[3] Friedrich Esch[1,2]*

[1] Chair of Physical Chemistry, Department of Chemistry, Technical University of Munich, Lichtenbergstr. 4, 85748 Garching, Germany

[2] Catalysis Research Center, Technical University of Munich, Lichtenbergstr. 4, 85748 Garching, Germany

[3] Dipartimento di Scienza dei Materiali, University of Milano-Bicocca, via Roberto Cozzi 55, 20125 Milano, Italy

[4] Institute of Physics, Ecole Polytechnique Fédérale de Lausanne, CH-1015 Lausanne, Switzerland

* bajlechner@tum.de

## S1. Pulsed reactivity data evaluation

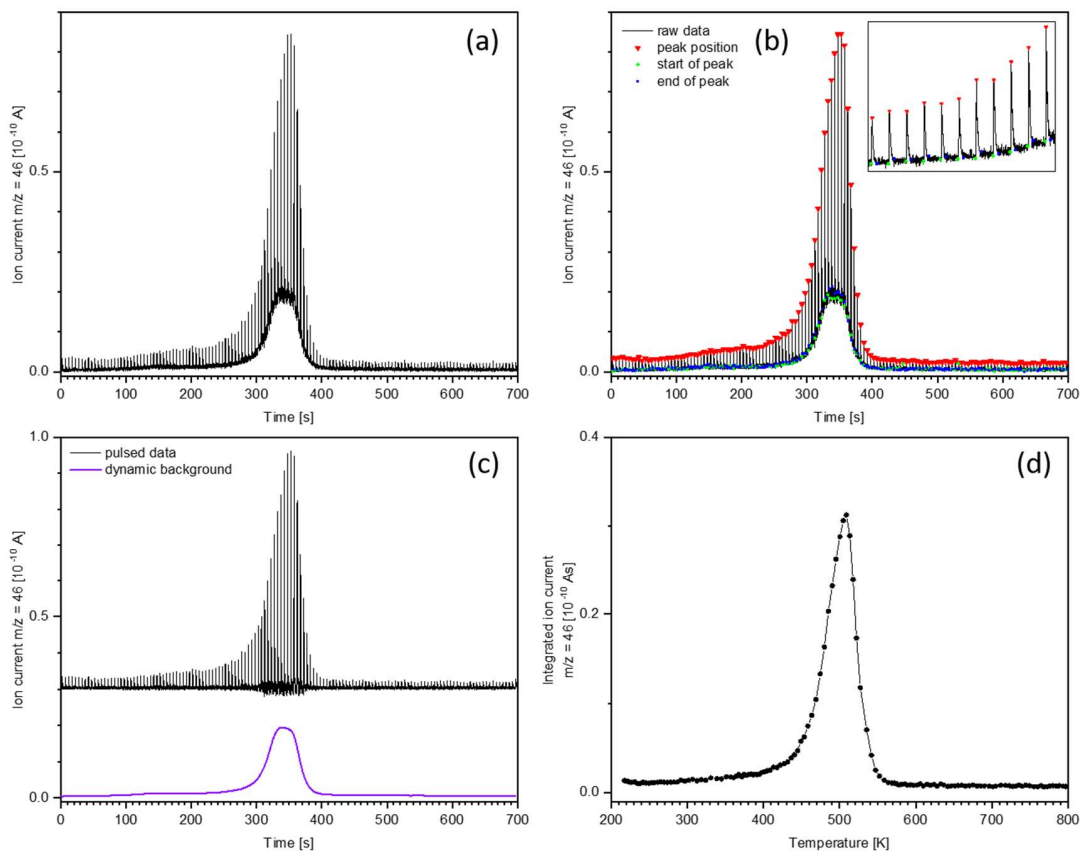

**Figure S1.** Evaluation procedure for pulsed valve sniffer measurements. (a) displays the measured raw data, consisting of a series of pulses convoluted with a continuous dynamic background (usually caused by desorption or reaction products). As both components contain information about the investigated system, it is necessary to separate them. First, the peak position (red), as well as the start (green) and end (blue) of each peak are detected, as shown in (b). The inset shows a zoomed-in section to better illustrate the method. The data in between the pulses (determined by the start and end points of the pulses) is smoothed and interpolated to determine the dynamic background, which is then subtracted. (c) shows the pulsed data after background subtraction (black, with an offset of $3 \times 10^{-11}$ A), as well as the subtracted background (purple). Finally, each individual pulse is integrated numerically and the resulting values are plotted vs. temperature, as shown in (d).

## S2. Characterization of magnetite bulk and (001) surface by DFT calculations

We started from the experimental crystal structure of $Fe_3O_4$ using a supercell with a $Fe_{24}O_{32}$ (($Fe_3O_4)_8$) formula. $Fe_3O_4$ (magnetite) has the structure of a cubic inverse spinel with $Fe^{3+}$ in the tetrahedral sites and a 50:50 mixture of $Fe^{2+}$ and $Fe^{3+}$ in the octahedral sites.[1] Tetrahedral and octahedral sublattices are anti-ferrimagnetically aligned in $Fe_3O_4$, such that the magnetic moments of the $Fe^{3+}$ cations on each sublattice cancel each other, and a nominal net magnetization of 4 $\mu_B$ per $Fe_3O_4$ formula unit derives from the $Fe^{2+}$ cations. We first optimized the lattice and internal coordinates with PBE+U+D3. The calculated lattice parameters (Table S1) show a slight distortion from the cubic symmetry. The lattice volume is overestimated by 2%, indicating a reasonable agreement of the hereby adopted computational method with the experiment.

**Table S1.** Calculated and experimental lattice parameters (a, b and c, Å) and cell volume (Å$^3$) of cubic $Fe_3O_4$.

|  | Calc. | Exp.[a] |
|---|---|---|
| **a** | 8.451 | 8.397 |
| **b** | 8.458 | 8.397 |
| **c** | 8.458 | 8.397 |
| **Cell Volume** | 604.59 | 592.1 |

[a] Data from ref [2]

The calculated electronic structure displays 4 unpaired electrons per formula unit, as reported in experiments (32 unpaired electrons in the unit cell). Figure S2 shows the spin density iso-surface of magnetite. Fe atoms in tetrahedral positions ($Fe^{3+}$) are spin down while the Fe atoms in octahedral sites are spin up (half $Fe^{3+}$ and half $Fe^{2+}$), consistent with an anti-ferrimagnetic ordering. In agreement with previous reports adopting a similar level of theory, magnetite is a semimetal, displaying conducting character on one spin channel and a small band gap ($\approx$ 0.5 eV) on the other.

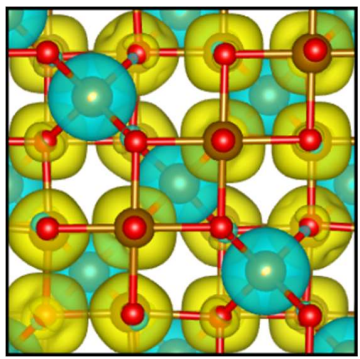

**Figure S2.** Spin-density iso-surface (0.01 |e|/Å$^3$) of bulk Fe$_3$O$_4$, with α (spin-up) electrons density in yellow and β (spin-down) in cyan.

The magnetite(001) surface is modelled following the sub-surface cation vacancy structure proposed by Parkinson.[3] The 1×1 cell's content is Fe$_{35}$O$_{48}$, where one Fe ion in an octahedral site of the subsurface layer has been removed. The hereby adopted slab model contains 12 atomic layers; the ions from the bottom 4 layers are kept frozen in their bulk lattice position, and all others are relaxed. As previously observed,[1] one Fe atom in the second atomic layer moves during the relaxation from an octahedral to a tetrahedral site, Figure S3. There are thus 13 Fe atoms in tetrahedral and 22 Fe atoms in octahedral positions. The total number of unpaired electrons in the supercell is 40 (3.3 per formula unit, showing a remarkable quenching of the magnetic moment with respect to the bulk).

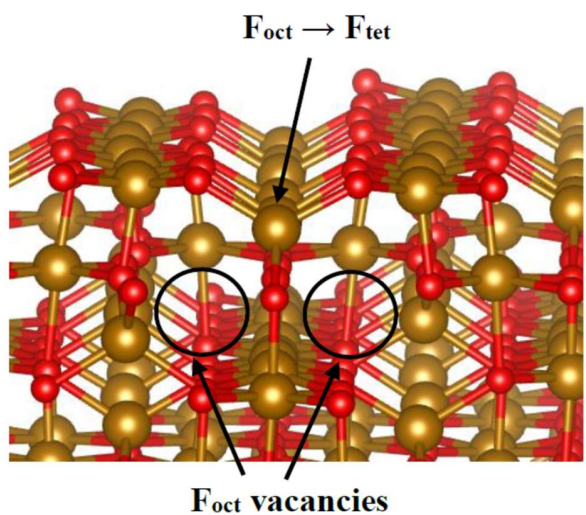

**Figure S3.** Optimized structure of the subsurface cation vacancy model of Fe$_3$O$_4$(001).

## S3. Formation of a surface oxygen vacancy on various sites

In this section we describe the effect of generating an oxygen vacancy on the surface of magnetite. Given the focus of the present paper on the surface properties of magnetite, we did not consider the vacancy formation in the subsurface and bulk regions. As shown in Figure S4, there are three non-equivalent O atoms on the first atomic layer of the $Fe_3O_4$(001) slab: O1 and O2 are next to an iron atom in a tetrahedral site and O3 is above an iron atom in an octahedral site (Figure S4). The formation energy of an oxygen vacancy ($E_f$), Table S2, is calculated as follows:

$$E_f = E[Fe_3O_{4-x}] + E[½\ O_2] - E[Fe_3O_4] \quad (1)$$

$$E'_f = E[Fe_3O_{4-x}] + E[O] - E[Fe_3O_4] \quad (2)$$

The O3 vacancy has the lowest formation energy of 2.91 eV; the other oxygens are removed at a somewhat higher cost, as shown in Table S2. Notably, the formation of a vacancy in O3 increases the net magnetization per formula unit from 3.4 $\mu_B$ (pristine surface) to 3.8 $\mu_B$, while the removal of O1 or O2 species has little effect on the magnetization.

**Table S2.** Formation energy of an oxygen vacancy ($E_f$ computed with respect to ½ $O_2$ and $E'_f$ computed with respect to atomic O, in eV) and magnetic moment per unit cell ($M_{total}$, $\mu_B$) on the (001) surface of $Fe_3O_4$, illustrated in Figure S4.

|  | $E_f$ | $E'_f$ | $M_{total}$ |
|---|---|---|---|
| $V_{O1}$ | 3.55 | 6.93 | 3.34 |
| $V_{O2}$ | 3.10 | 6.48 | 3.32 |
| $V_{O3}$ | 2.91 | 6.30 | 3.79 |

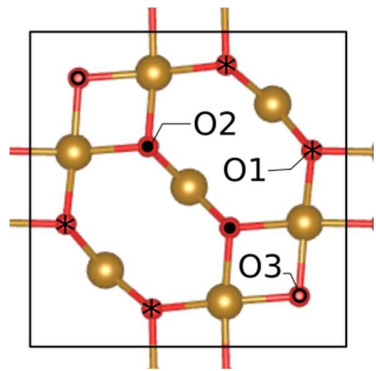

**Figure S4**. Top view of the first and second atomic layers of the $Fe_3O_4$(001) unit cell. O1, O2 and O3 indicate the non-equivalent O atoms on the surface of the $Fe_3O_4$(001) slab.

**S4. CO desorption from bare magnetite and magnetite-supported clusters in TPD**

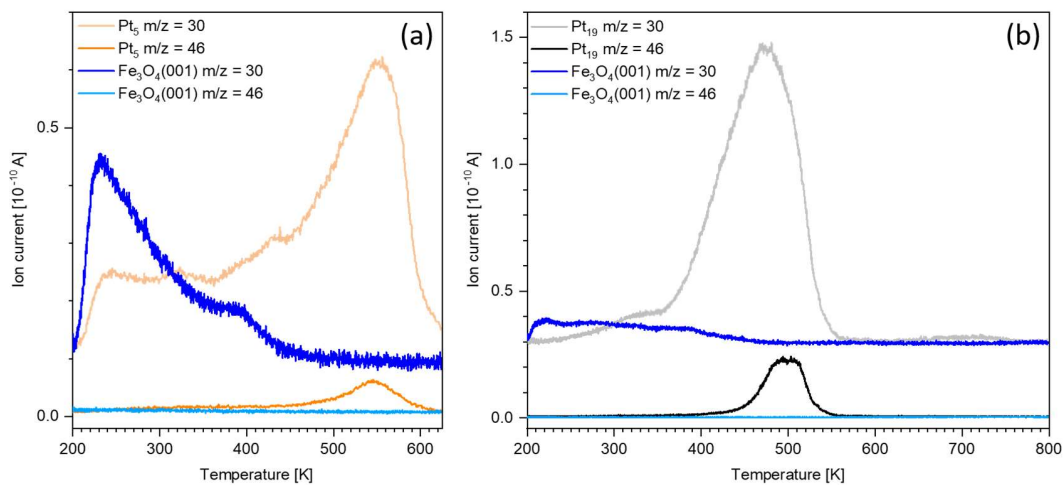

**Figure S5.** Saturation $C^{18}O$ TPD curves of (a) $Pt_5$ (orange) and (b) $Pt_{19}$ (gray) supported on $Fe_3O_4(001)$ compared to the corresponding desorption curves from the clean support (blue). The heating rate was 1 K/s. The $C^{18}O$ (m/z = 30) and $C^{18}O^{16}O$ (m/z = 46) signals are shown. The clean magnetite surface exhibits several overlapping, not very distinct CO desorption features in the temperature region investigated here. The desorption starts immediately upon heating and is finished below 450 K. These background desorption features are the high temperature shoulder of a much larger desorption peak located around 180 K, which is attributed to an unidentified magnetite surface defect.[4] It can be observed that especially the beginning of the background CO desorption around 230 K is much less pronounced with clusters deposited on the $Fe_3O_4(001)$ surface, indicating that a part of the defects are no longer accessible CO adsorption sites. This effect can be explained by at least some of the clusters occupying surface defect sites. Furthermore, it is shown that no $CO_2$ desorption or production from the bare $Fe_3O_4(001)$ surface is detectable in the measured temperature region (cyan).

## S5. Ripening of Pt₅ in STM

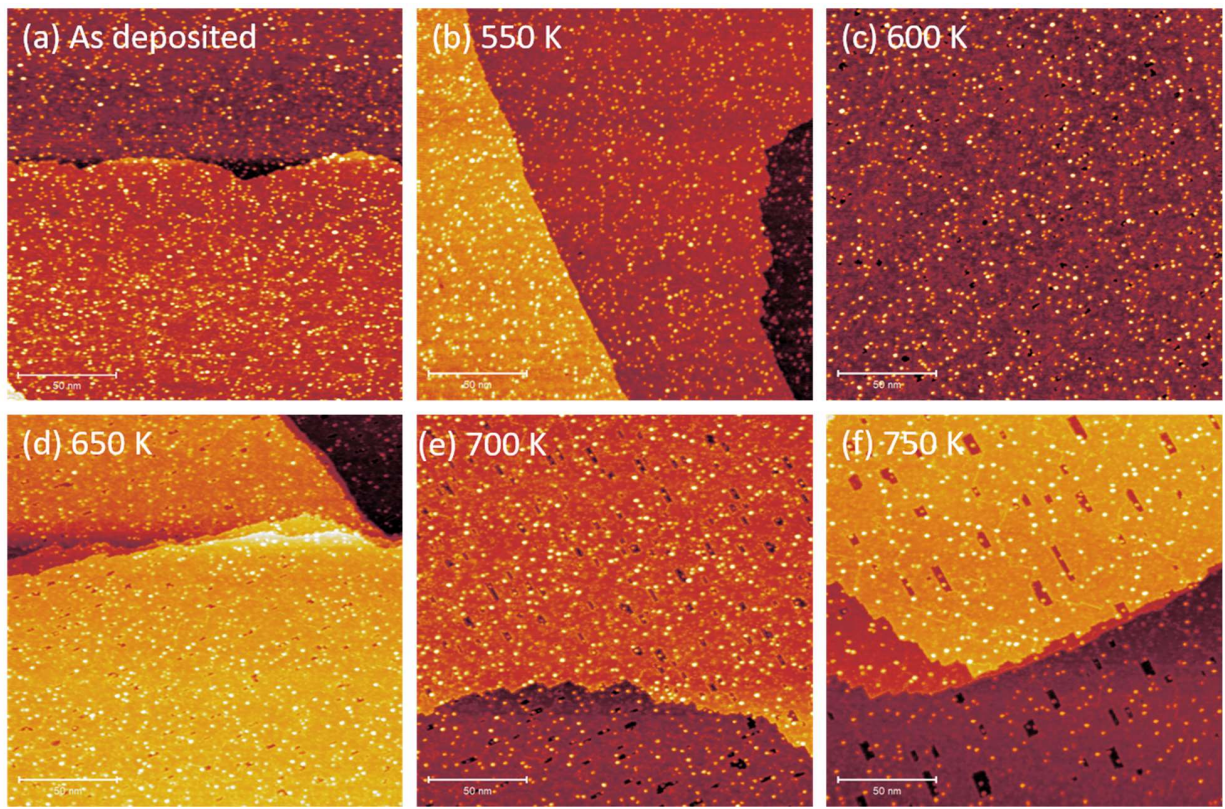

**Figure S6.** STM image series of Pt$_5$ clusters (0.05 clusters/nm$^2$) on Fe$_3$O$_4$(001), (a) as deposited at RT, (b-f) measured at RT after annealing to the temperatures indicated, respectively. Up to an annealing temperature of 600 K, the number of clusters stays constant. Above 650 K, gradual cluster ripening is observed, yielding fewer clusters with an average size of about Pt$_{15-20}$ at 750 K. The approximate size is calculated by comparing the apparent height and number of clusters at 750 K with the as-deposited sample. Upon annealing, hole formation around some clusters is observed, due to lattice oxygen migrating onto the clusters and reaction with background CO, while the residual, reduced iron atoms diffuse into the bulk. *Imaging parameters: $V_b$ = 1.50 V; $I_t$ = 300 pA.*

## S6. Alternating CO and $O_2$ pulses on $Pt_5$/$Fe_3O_4$(001)

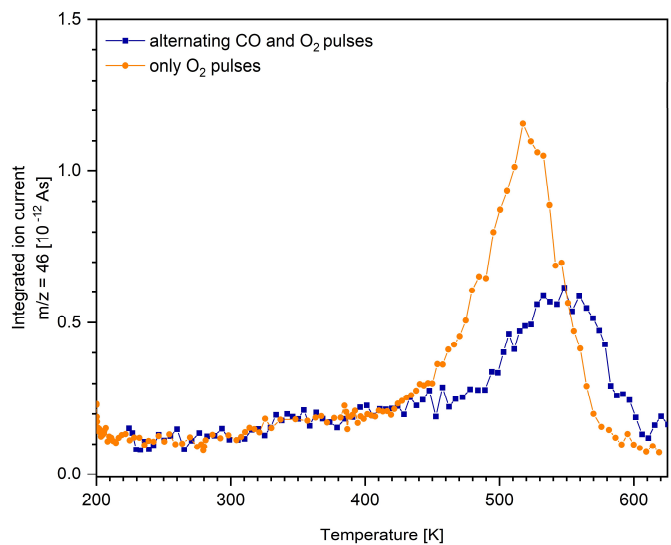

**Figure S7.** $CO_2$ production synchronized with $O_2$ pulses as a function of temperature, obtained by pulsing only $O_2$ (orange) and alternatingly pulsing CO and $O_2$ (blue) on CO pre-covered $Pt_5$ clusters on $Fe_3O_4$(001). Both curves have been normalized to the Pt atom coverage. When additionally pulsing CO, the onset of the $CO_2$ production peak synchronized with $O_2$ pulses is shifted to higher temperatures, indicating a stronger initial poisoning of the cluster surface by CO since the CO pulses shift the desorption equilibrium. The peak position is similarly shifted from 520 K to 550 K. This indicates a reaction limitation due to a lack of CO when only pulsing $O_2$.

## S7. Loss of CO adsorption capability upon cluster encapsulation

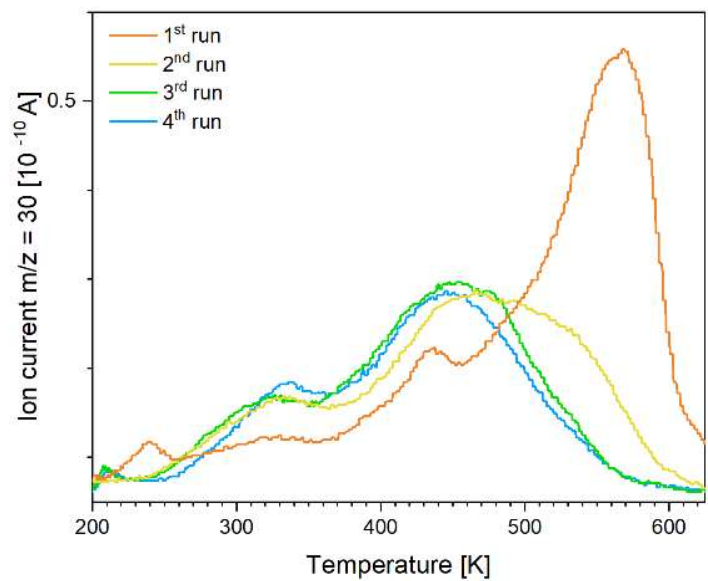

**Figure S8.** Subsequent saturation $C^{18}O$ TPD curves (m/z = 30) of $Pt_5$ clusters (0.05 clusters/nm$^2$) on $Fe_3O_4$(001), corresponding to the TPD experiments shown in Figure 2. For every run, the surface was saturated with $C^{18}O$ at 200 K, the heating rate was 1 K/s. The main peak in the first run (orange) at 550 K is attributed to the clusters. Several lower temperature features are observed as well, which most certainly are related to surface defects (as described in section 3.1 in the main text). In the second run (yellow) the cluster-related desorption feature decreases significantly, while two broad lower temperature peaks arise at 325 K and 450 K. In the third TPD run (green), the CO desorption from the cluster vanishes almost completely, while the lower temperature peaks become slightly more pronounced. In the fourth run (blue), only the lower temperature peaks occur. This subsequent decrease in CO adsorption capability of the clusters is attributed to encapsulation as a result of SMSI, comparable to Pt nanoparticles supported on $Fe_3O_4$(001) that become encapsulated by an FeO layer upon annealing.[5] The TPD series indicates that after the first run, the clusters are partially encapsulated, with still a fraction of their surface accessible for CO adsorption. In the following runs, the clusters become completely encapsulated, thus no cluster-related CO desorption is observed anymore. The lower temperature features appear simultaneously with the encapsulation of the clusters and may therefore originate from CO desorption from the encapsulating layer, which is expected to be non-stoichiometric, reduced iron oxide, since the lower temperature peaks are well within the temperature range of CO desorption from FeO surfaces.[6] However, more experiments have to be done for further clarification.

## S8. Hole formation in cluster vicinity

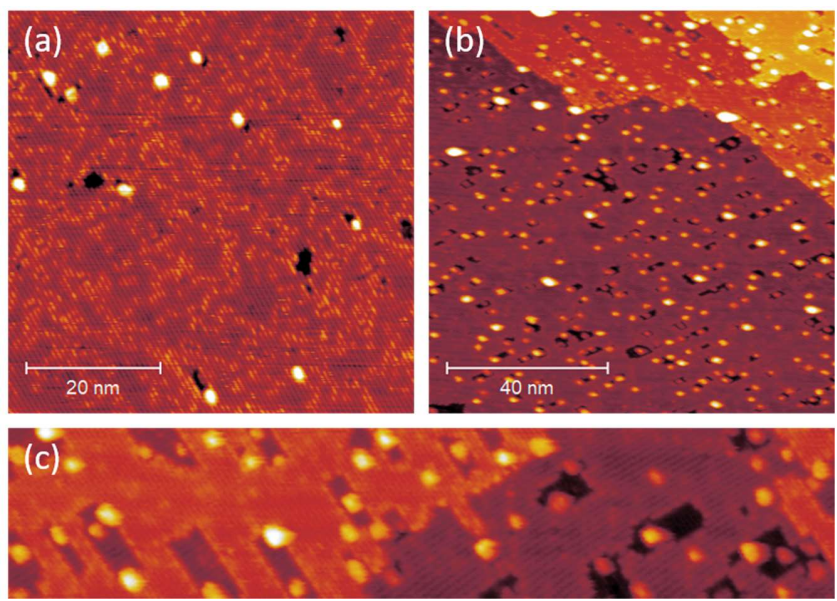

**Figure S9.** Hole formation around Pt clusters on $Fe_3O_4(001)$. (a) shows $Pt_{10}$ clusters (0.01 clusters/nm$^2$) measured at 573 K. Clear hole formation around the cluster perimeter can be observed caused by the removal of lattice oxygen around the clusters by reverse spillover and reaction with CO from the chamber background, followed by Fe migration into the bulk. (b) and (c) display $Pt_{19}$ (0.05 clusters/nm$^2$) measured at RT after annealing to 590 K in 1 x 10$^{-7}$ mbar CO for 2 minutes. Again, hole formation around the clusters can be observed, but much more pronounced compared to samples annealed in UHV as a consequence of the higher CO partial pressure. The holes exhibit a rectangular shape which changes direction between two neighboring terraces, following the direction of the atomic rows of the support. This becomes very obvious in the zoomed-in area in (c). The holes are fairly large compared to the clusters, indicating a facile oxygen diffusion towards the clusters at elevated temperatures. The edges of the holes are straight and terminated by an iron row of the magnetite lattice, but more rough perpendicular to them. This indicates that the oxygen diffusion is more favorable along an atomic row, suggesting the initial removal of an oxygen atom from a pristine atomic row to be most difficult. *Imaging parameters:* $V_b = 1.50$ V; (a) $I_t = 500$ pA, (b, c) $I_t = 300$ pA.

## S9. Platinum clusters in the gas phase: structure relaxation and oxygen adsorption

The minimum energy structures of gas phase Pt clusters have been studied in previous DFT investigations.[7–13] For instance, the calculations with a B3PW91 hybrid functional show that a $Pt_5$ cluster with distorted squared pyramid geometry, Figure S10(iii), is the lowest energy isomer.[7] Here, we have considered six of the most stable isomers, and recomputed their stabilities and structures at the PBE+U+D3 level, Figure S10. Four isomers are within an energy of 200 meV, and all six isomers are within 0.35 eV, with energy differences of 70-80 meV, respectively. This suggests a high fluxionality in the gas phase. Of course, once deposited on an oxide support, the structural flexibility of the Pt clusters could be reduced.

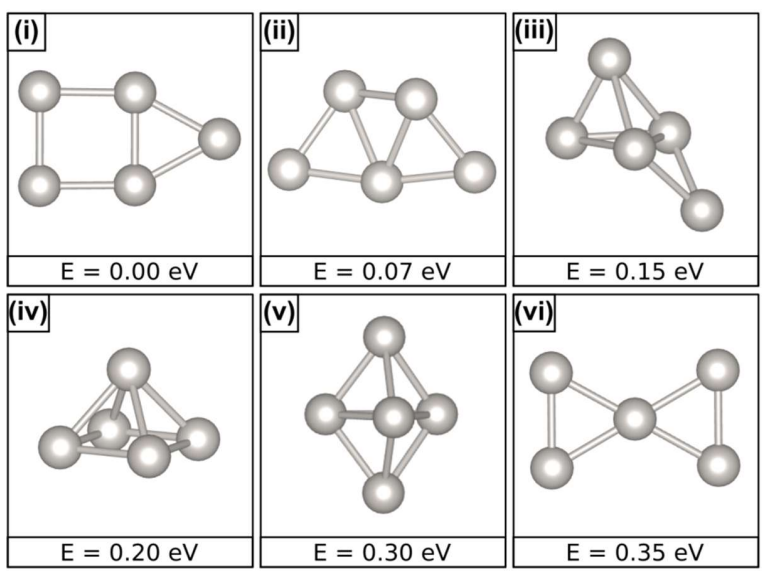

**Figure S10.** Optimized structures and relative energies of isomers of the $Pt_5$ gas phase cluster.

We consider the preferred adsorption sites for a single O atom on the various isomers of gas phase $Pt_5$. Altogether, we have considered 25 possible structures, and this is not necessarily an exhaustive sample. The adsorption of an O atom on the (i) to (vi) $Pt_5$ clusters, summarized in Table S3, occurs in three positions: terminal, bridge and hollow. In many cases, the position of the adsorbed O atom and/or the shape of cluster change during optimization. Table S3 reports the optimized structures, their relative energy ($E_R$), and the adsorption energy ($E_{ADS}$) of an O atom with respect to the most stable $Pt_5$ (i) cluster and ½ $O_2$ molecule, calculated as follows:

$Pt_5(i) + ½ O_2 \rightarrow$ O-$Pt_5$ (n)    $\Delta E = E_{ADS}$ (3)

The most stable gas phase Pt$_5$O cluster is planar, with the Pt atoms forming three fused triangles and the O atom adsorbed in a Pt-Pt bridge position. The next most stable isomer has a completely different structure: it is three-dimensional with the adsorbed O in a terminal position, bound to an apical Pt atom. It is 0.42 eV higher in energy than the ground state structure. Also the next two stable isomers 3 and 4, Table S3, show the presence of O in a terminal position. In general, an O atom binds preferentially in a Pt-Pt bridge site only if the Pt-Pt distance is such to favour its coordination. Cases where the O atom is in a hollow site are clearly higher in energy.

**Table S3.** Relative energy ($E_R$) and adsorption energy ($E_{ADS}$) of an adsorbed O atom on Pt$_5$ clusters.

| Structure | | $E_R$, eV | $E_{ADS}$, eV | Structure | | $E_R$, eV | $E_{ADS}$, eV |
|---|---|---|---|---|---|---|---|
| 1 | 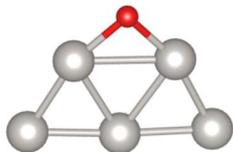 | 0.00 | -2.42 | 13 | 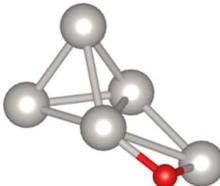 | 0.95 | -1.47 |
| 2 | 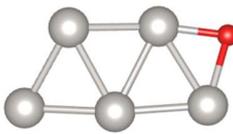 | 0.42 | -2.00 | 14 | 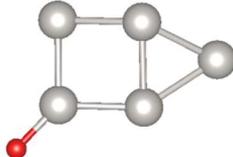 | 0.98 | -1.44 |
| 3 | 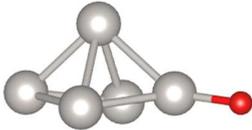 | 0.53 | -1.89 | 15 | 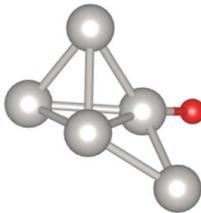 | 1.09 | -1.33 |
| 4 | 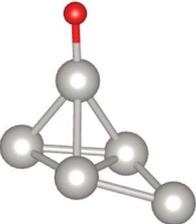 | 0.63 | -1.80 | 16 | 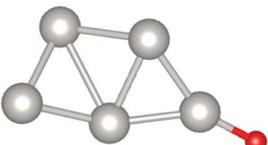 | 1.11 | -1.31 |
| 5 | 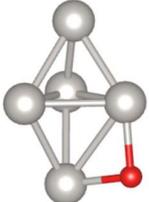 | 0.65 | -1.78 | 17 | 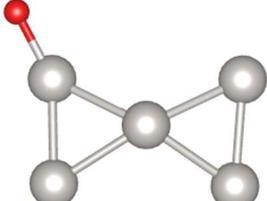 | 1.16 | -1.27 |
| 6 | 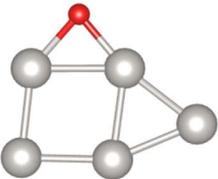 | 0.67 | -1.75 | 18 | 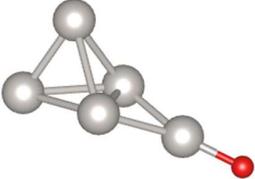 | 1.18 | -1.24 |
| 7 | 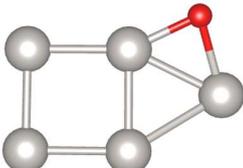 | 0.67 | -1.75 | 19 | 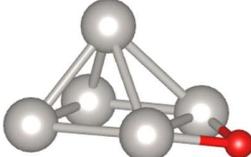 | 1.21 | -1.21 |
| 8 | 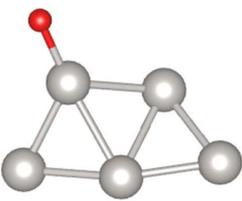 | 0.77 | -1.65 | 20 | 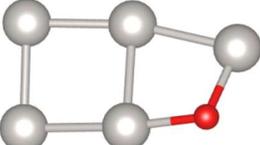 | 1.26 | -1.17 |
| 9 | 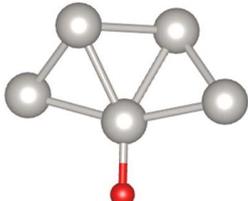 | 0.83 | -1.60 | 21 | 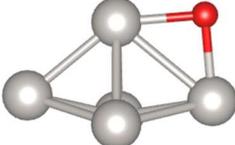 | 1.30 | -1.12 |

| 10 | 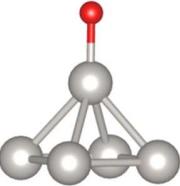 | 0.83 | -1.59 | 22 | 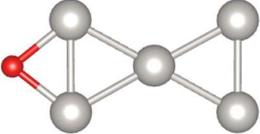 | 2.11 | -0.31 |
| 11 | 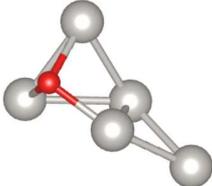 | 0.86 | -1.56 | 23 | 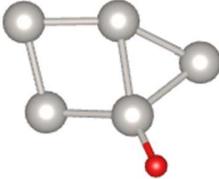 | 2.12 | -0.30 |
| 12 | 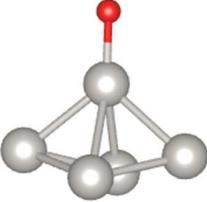 | 0.91 | -1.52 | 24 | 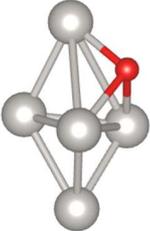 | 2.18 | -0.24 |
| | | | | 25 | 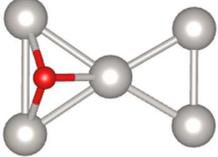 | 2.22 | -0.20 |

## S10. Adsorption of Pt$_5$ on Fe$_3$O$_4$(001) and formation of oxygen vacancies

**Table S4.** Adsorption energy (E$_{ADS}$, eV), Bader charge (q, |e|), and spin polarization of adsorbed Pt$_5$ clusters on Fe$_3$O$_4$ and total spin magnetic moment per unit cell (M$_{total}$, µ$_B$).

| Figure | E$_{ADS}$[a] | q(Pt$_5$) | M (Pt$_5$) | M$_{total}$ |
|---|---|---|---|---|
| 4(a) | -4.51 | 0.33 | 0.08 | 2.59 |
| 4(b) | -3.01 | 0.25 | 1.12 | 3.17 |
| 4(c) | -2.73 | 0.33 | 0.19 | 3.22 |

[a] Energy change related to the reaction Pt$_5$(i) + Fe$_3$O$_4$ → Pt$_5$(n)/Fe$_3$O$_4$

The Pt$_5$ adsorption energies, reported in Table S4, are calculated with respect to the clean support and the gas phase cluster in its most stable configuration (capped square, Figure S10(i)).

The formation energy of the oxygen vacancy (E$_f$), Table S5, has been calculated according to the following equation:

$$E_f = E[(Pt_5/Fe_3O_{4-x})] + E[½\, O_2] - E[(Pt_5/Fe_3O_4)] \qquad (4)$$

**Table S5.** O vacancy energy (E$_f$, eV), Bader charge (q, |e|), spin polarization of an adsorbed Pt$_5$ cluster (Figure 4(a)) and total spin magnetic moment per unit cell (M, µ$_B$). The position of the O atom removed is shown in the Figure.

| | E$_f$ | q(Pt$_5$) | M (Pt$_5$) | M$_{total}$ |
|---|---|---|---|---|
| V$_{O1}$ | 3.14 | -0.27 | 0.20 | 2.61 |
| V$_{O2}$ | 1.98 | -0.37 | 0.27 | 3.24 |
| V$_{O3}$ | 2.66 | -0.50 | 0.28 | 2.49 |
| V$_{O4}$ | 2.46 | 0.24 | 0.35 | 3.40 |

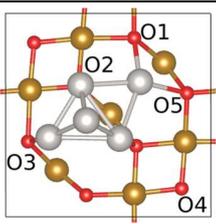

As discussed in the main text, we observe a general decrease in the formation energy in the presence of the platinum clusters with respect to the clean support. The reduction of the substrate upon oxygen removal, however, does not imply a remarkable charge transfer to the cluster, as shown by the Bader charges. In some cases, in particular V$_{O1}$ and V$_{O3}$, a significant decrease of the net magnetization is reported.

## S11. Lattice oxygen reverse spillover

*(i) Displacement of one oxygen atom*

The study of the formation of an O vacancy on the clean support has shown that O2 is the easiest O to remove, as shown in Table S5. In this section, we consider the energetic cost of displacing the O2 atom from the support and adsorbing it on a $Pt_5$ cluster. The process corresponds to the final state of a lattice oxygen reverse spillover effect. The O atom has been re-adsorbed on various sites of the supported $Pt_5$ cluster, always starting from the most stable structure, Figure 4(a) in the main text, corresponding to the $Pt_5$(iii) gas phase isomer. Other structures have also been considered, some of them being unstable. For instance, O adsorption on one of the triangular facets of $Pt_5$ always resulted in an O atom bound to a bridge site. The most stable isomers found are shown in Figure S11 in order of stability (see also Table S6).

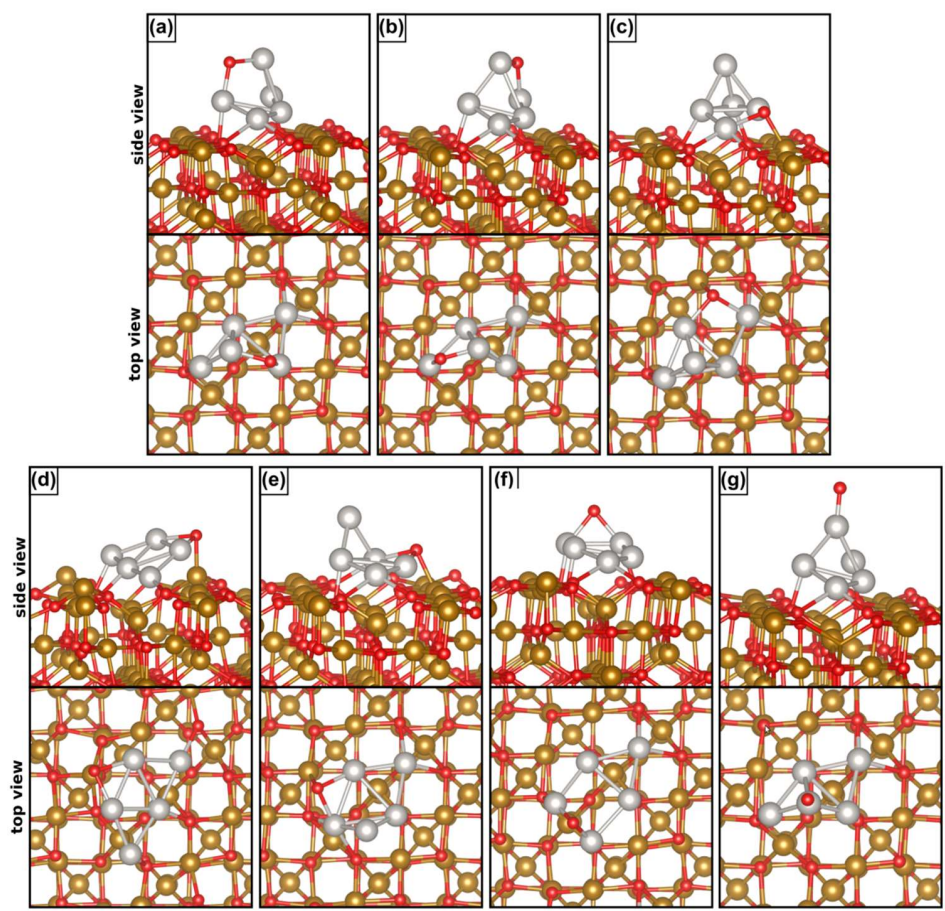

**Figure S11.** Various isomers of $Pt_5O$ supported on $Fe_3O_4$. The structures have been obtained starting the geometry optimization from $Pt_5/Fe_3O_4$, Figure 4(a).

The most stable structure is that of a capped rhombus $Pt_5$ with an O atom adsorbed on one bridge site, shown in Figure S11(a). The corresponding energy is -0.40 eV lower than that of $Pt_5/Fe_3O_5$ (see Table S6).

**Table S6.** O reverse spillover energy ($E_{SO}$, eV), Bader charge (q, |e|), spin polarization of $Pt_5/Fe_3O_4(001)$ and the total spin magnetic moment per unit cell ($M_{total}$, $\mu_B$).

| Figure | $E_{SO}$[a] | q($Pt_5$) | q($O_{spill}$) | M ($Pt_5$) | $M_{total}$ |
|---|---|---|---|---|---|
| S10 (a) | -0.40 | 0.25 | -0.68 | 0.03 | 3.26 |
| S10 (b) | -0.15 | 0.29 | -0.68 | 0.70 | 3.32 |
| S10 (c) | -0.04 | 0.67 | -0.86 | 0.48 | 3.27 |
| S10 (d) | 0.40 | 0.14 | -0.82 | 0.98 | 3.80 |
| S10 (e) | 0.74 | 0.27 | -0.81 | 0.08 | 4.51 |
| S10 (f) | 0.86 | 0.40 | -0.72 | 0.07 | 3.92 |
| S10 (g) | 1.68 | 0.22 | -0.57 | 0.13 | 3.46 |

[a] $E_{SO}$ refers to the energy of this reaction: $Pt_5/Fe_3O_4 \rightarrow O\text{-}Pt_5/Fe_3O_{4-x}$

This shows that the lattice oxygen reverse spillover in this specific case is an exothermic process. A second isomer, Figure S11(b), is only 0.25 eV higher in energy and differs from the isomer of Figure S11(a) only for the Pt-Pt bond that is bridge-bonded by O. A third interesting structure is shown in Figure S11(c). Here, the O atom is bridging a Pt-Pt bond at the cluster/oxide interface. It can be considered the first step in the migration of an O atom from the support onto the cluster. The isomer shown in Figure S11(c) is -0.04 eV lower in energy than the starting structure, i.e. is thermoneutral with respect to the case where no O vacancy has been formed on the support, and no oxygen has been transferred to $Pt_5$. Other $Pt_5O$ isomers are shown in Figure S11(d)-(g); they are all higher in energy than the three best structures shown in Figure S11(a)-(c).

*(ii) Displacement of two oxygen atoms*

Next, we have considered the spillover of two oxygen atoms from the magnetite surface onto $Pt_5$. Different O atoms have been removed from the support, and different adsorption positions have been considered on $Pt_5O$, starting from the most stable structures obtained for the case of a single O reverse spillover, as shown Figure S11. The results are reported in Figure S12 and Table S7.

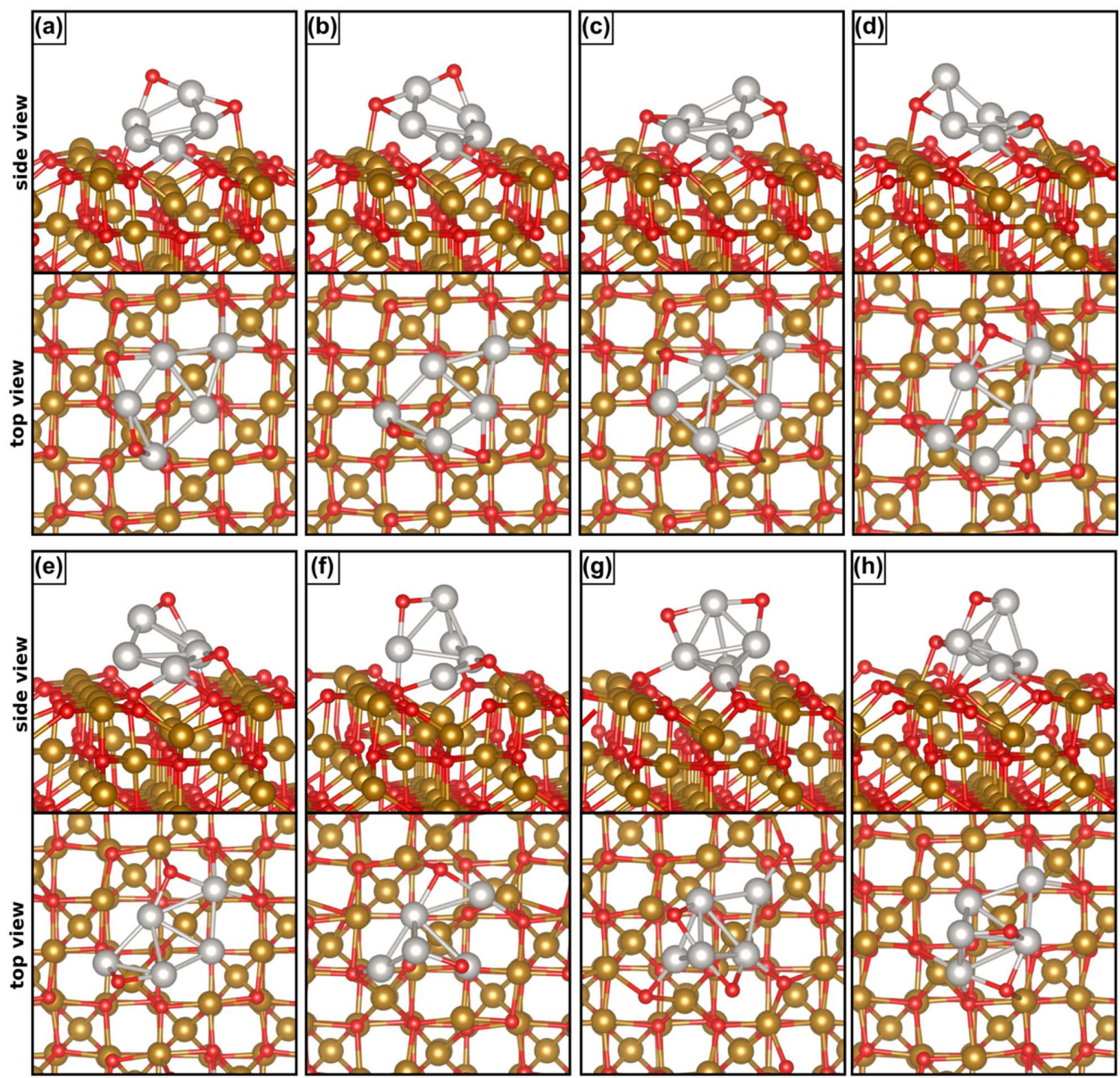

**Figure S12.** Various isomers of $Pt_5O_2$ supported on $Fe_3O_{4-x}$. The structures have been obtained starting the geometry optimization from $Pt_5O/Fe_3O_{4-x}$, Figure 5(a). (a) to (d) O2 and O4 spillover; (e) O1 and O2 spillover; (g) and (h) O2 and O4 spillover (see Table S5).

We found two isomers, shown in Figure S12(a) and (b) whose structures are *lower* in energy, by -0.64 and by -0.56 eV, Table S7, than the regular, non-defective surface with adsorbed $Pt_5$. This is an important result which shows that multiple oxygen transfer from the support to a small Pt cluster is thermodynamically favorable, and by a considerable amount. These two structures have been obtained by removing the O2 and O4 atoms of the surface, just under the $Pt_5$ cluster, as shown in Figure 5(b) in the main text. An important observation is that after geometry optimization, the $Pt_5O_2$ cluster changes its structure and from 3D (Figure S11(a)) it becomes nearly flat, as shown in Figure S12(a) and (b).

$Pt_5$ structures that maintain the original 3D shape, a capped rhombus (Figures S10(f) and (g)), are slightly below or slightly above the $Pt_5/Fe_3O_4$ reference, but clearly higher in energy than the nearly 2D ground state structures.

These results show two important trends: (1) multiple lattice oxygen reverse spillover is an energetically favorable process; and (2) the adsorption of two oxygen atoms on $Pt_5$ leads to a restructuring that tends to flattens the cluster structure. In order to confirm these trends, in the next Section we considered the spillover of three oxygen atoms.

**Table S7.** Oxygen spillover energy ($E_{SO}$, eV), Bader charge (q, |e|), spin polarization of $Pt_5/Fe_3O_4(001)$ and the total spin magnetic moment per unit cell (M, $\mu_B$).

|         | Figure  | $E_{SO}^{(a)}$ | q($Pt_5$) | q ($2O_{spill}$) | M ($Pt_5$) | $M_{total}$ |
|---------|---------|-------|---------|-------------|---------|---------|
| O2, O4  | S11 (a) | -0.64 | 0.37    | -1.52       | 0.21    | 2.54    |
|         | S11 (b) | -0.56 | 0.48    | -1.50       | 0.15    | 3.92    |
|         | S11 (c) | 0.59  | 0.41    | -1.62       | 0.19    | 3.96    |
|         | S11 (d) | 0.83  | 0.46    | -1.61       | 0.19    | 3.93    |
|         | S11 (e) | 0.86  | 0.49    | -1.52       | 0.56    | 2.54    |
| O1, O2  | S11 (f) | 0.30  | 0.53    | -1.58       | 0.65    | 3.34    |
| O2, O4  | S11 (g) | -0.04 | 0.73    | -1.40       | 0.35    | 3.18    |
|         | S11 (h) | 1.66  | 1.07    | -1.46       | 0.40    | 3.99    |

(a) $E_{SO}$ refers to the process: $Pt_5/Fe_3O_4 \rightarrow 2O\text{-}Pt_5/Fe_3O_{4-x}$

*(iii) Displacement of three oxygen atoms*

We finally considered the case where three oxygens are displaced from the magnetite surface onto the $Pt_5$ cluster, with formation of a $Pt_5O_3$ unit and three oxygen vacancies. The number of potential isomers is huge, and we restricted the analysis to the sites where oxygen is easier to remove, O1, O2, and O4 or O1, O2 and O5, and to a few structures derived from the best isomers of $Pt_5O_2$, as shown in Figure S13 and Table S8. The two most stable structures are shown in Figure S13 (a) and (b) and correspond to a flat or nearly flat $Pt_5$ with three oxygen atoms in bridge sites; some of them remain coordinated to Fe atoms of the support, thus providing anchoring points for the nanocluster. These two structures are only 0.17 and 0.23 eV higher in energy, respectively, than the pristine $Pt_5/Fe_3O_4$, showing that even the transfer of three oxygen atoms is energetically possible (the reaction is only slightly endothermic). We cannot exclude that some more favorable isomers exist, but this already provides sufficient evidence that multiple oxygen transfer from magnetite to supported Pt clusters is possible.

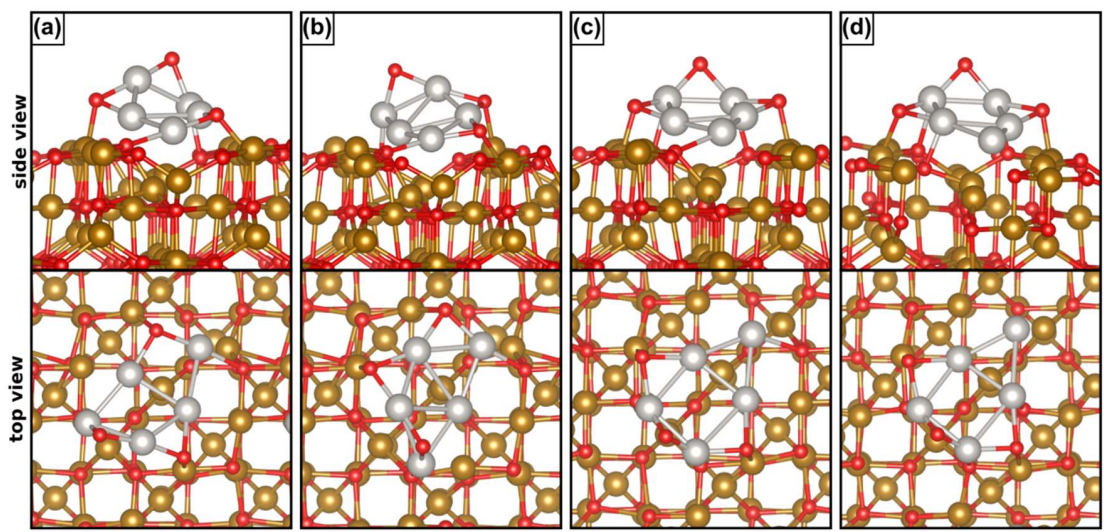

**Figure S13.** Various isomers of $Pt_5O_3$ supported on $Fe_3O_{4-x}$. The structures have been obtained starting the geometry optimization from $Pt_5O_2/Fe_3O_{4-x}$, (a) to (c) O1, O2 and O4 spillover; (d) O1, O2 and O5 spillover (see Table S5).

**Table S8.** Three O spillover energy ($E_{SO}$, eV), Bader charge (q, |e|), spin polarization of $Pt_5/Fe_3O_4(001)$ and the total spin magnetic moment per unit cell (M, $\mu_B$).

| Figure | $E_{SO}$[a] | q(Pt$_5$) | q (3O$_{spill}$) | M (Pt$_5$) | M$_{total}$ |
|---|---|---|---|---|---|
| S12 (a) | 0.17 | 0.85 | -2.43 | 0.31 | 3.16 |
| S12 (b) | 0.23 | 1.00 | -2.48 | 0.10 | 4.58 |
| S12 (c) | 0.95 | 0.71 | -2.35 | 0.79 | 2.57 |
| S12 (d) | 3.85 | 0.93 | -2.68 | 0.11 | 3.33 |

[a] $E_{SO}$ refers to the process: $Pt_5/Fe_3O_4 \rightarrow 3O\text{-}Pt_5/Fe_3O_{4-x}$

## S12. Impact of lattice oxygen reverse spillover on cluster height

In Table S9 we provide some evidence that the lattice oxygen reverse spillover mechanism, with formation of $Pt_xO_y$ clusters, can contribute to a structural change and in particular to a transition from 3D to 2D. For all the structures considered, we have computed the average vertical distance of the Pt atoms from the surface of magnetite. Since the atoms of the surface may be involved in structural changes, we have taken the vertical distance of the Pt atoms from the bottom layer of our slab, whose coordinates are fixed. The most stable $Pt_5$ cluster, shown in Figure 4 (a) in the main text, is taken as reference. We define $\Delta Z$ as the difference of the average distance of the Pt atoms in the various oxidized clusters (see Figures S10 and S11).

The most stable $Pt_5O$ cluster (see Figure S11 (a)), exhibits $\Delta Z = -0.102$ Å; the most stable $Pt_5O_2$ cluster (see Figure S12 (a)) has $\Delta Z = -0.386$ Å; and the most stable $Pt_5O_3$ cluster (see Figure S13 (a)) has $\Delta Z = -0.426$ Å. This trend is not fortuitous and is found also for the other most stable $Pt_xO_y$ clusters, as shown in Table S9. The physical reason for this is that some of the O atoms bind at the cluster periphery, and interact also with Fe ions on the surface. In order to increase this bonding, the cluster flattens its structure, resulting in a transition from 3D to 2D.

**Table S9.** Different distances of $Pt_5O_x$ clusters to the magnetite support with respect to $Pt_5/Fe_3O_4$.

|  | Figures | $Z_{avg}$(Å) | $\Delta Z$(Å) |
|---|---|---|---|
| **$Pt_5/Fe_3O_4$** | **4 (a)** | **13.734** | **0.000** |
|  | 4 (b) | 13.908 | 0.174 |
|  | 4 (c) | 13.424 | -0.310 |
| One O spillover | S10 (a) | 13.632 | -0.102 |
|  | S10 (b) | 13.662 | -0.072 |
|  | S10 (c) | 13.735 | 0.001 |
|  | S10 (d) | 13.478 | -0.256 |
|  | S10 (e) | 13.419 | -0.315 |
|  | S10 (f) | 13.245 | -0.489 |
|  | S10 (g) | 13.631 | -0.103 |
| Two O spillover | S11 (a) | 13.348 | -0.386 |
|  | S11 (b) | 13.344 | -0.390 |
|  | S11 (c) | 13.408 | -0.326 |
|  | S11 (d) | 13.414 | -0.320 |
|  | S11 (e) | 13.284 | -0.450 |
|  | S11 (f) | 13.547 | -0.187 |
|  | S11 (g) | 13.512 | -0.222 |
|  | S11 (h) | 13.478 | -0.256 |
| Three O spillover | S12 (a) | 13.308 | -0.426 |
|  | S12 (b) | 13.435 | -0.299 |
|  | S12 (c) | 13.172 | -0.562 |
|  | S12 (d) | 13.312 | -0.062 |

$Z_{avg} = \Sigma\ z(Fe,O)\ \text{bottom layer}/m - \Sigma\ z(Pt)\ \text{cluster}/n$

$\Delta Z = Z_{avg}(Pt_xO_y) - Z_{avg}(Pt_5)$

## S13. Iron spillover

There are two kinds of non-equivalent Fe atoms on the first and second atomic layers of the $Fe_3O_4(001)$ slab: Fe atoms in the octahedral and tetrahedral positions, respectively, as shown in Figure S14. The formation energy of an Fe vacancy ($E_f$), listed in Table S10, is calculated according to the following equations:

$$E_f = E[Fe_{3-x}O_4] + E[Fe\ (s)] - E[Fe_3O_4] \quad (5)$$

$$E'_f = E[Fe_{3-x}O_4] + E[Fe\ (g)] - E[Fe_3O_4] \quad (6)$$

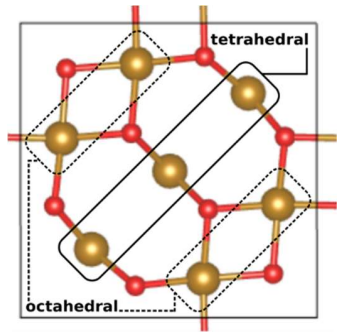

**Figure S14**. Top view of the first and second atomic layers of the $Fe_3O_4(001)$ unit cell. Fe atoms in octahedral and tetrahedral sites on the surface of an $Fe_3O_4(001)$ slab are marked by dashed and full lines, respectively.

**Table S10.** Formation energy of an iron vacancy ($E_f$ computed with respect to solid Fe and $E'_f$, computed with respect to atomic Fe, in eV) and magnetic moment per unit cell ($M_{total}$, $\mu_B$) on the (001) surface of $Fe_3O_4$.

|  | $E_f$ | $E'_f$ | $M_{total}$ |
|---|---|---|---|
| $Fe_{3-x}O_4$ ($V_{Fe\_tetrahedral}$) | 6.60 | 9.35 | 3.69 |
| $Fe_{3-x}O_4$ ($V_{Fe\_octahedral}$) | 6.64 | 9.39 | 2.08 |

Here, we discuss the Fe vacancy formation for the case of the most stable $Pt_5/Fe_3O_4$ structure (Figure 4(a)). As shown in Figure S15, there are 7 non-equivalent Fe atoms in the first and second atomic layers of the $Fe_3O_4(001)$ slab. We have considered two vacancies in tetrahedral (F1 and F2 shown in Figure S15) and two vacancies in octahedral (F3 and F4 shown in Figure S15) positions on the surface of the $Pt_5/Fe_3O_4$ structure. The optimized structures are shown in Figure S16.

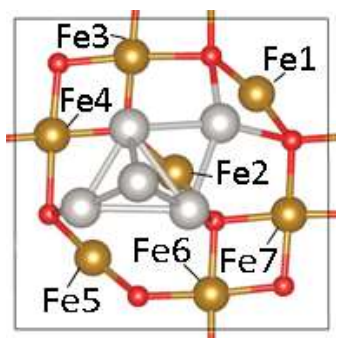

**Figure S15.** The non-equivalent Fe atoms in the first and second atomic layers of the $Pt_5/Fe_3O_4$ structure shown in Figure 4(a).

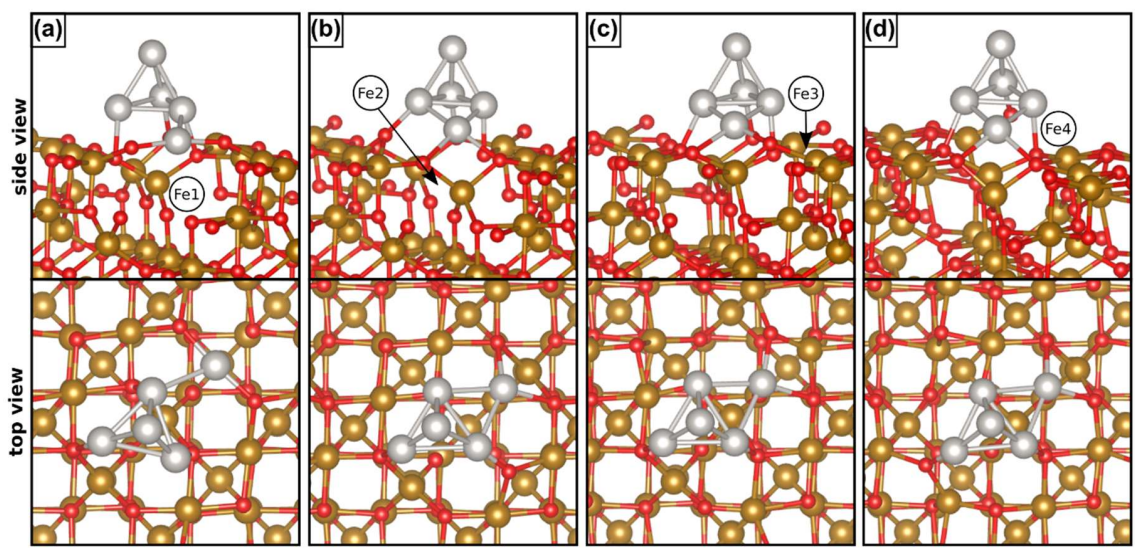

**Figure S16.** The optimized structures of Fe vacancies in the first layer of the $Pt_5/Fe_3O_4$ structure shown in Figure 4 (a).

The formation energy of the iron vacancy ($E_f$), given in Table S11, has been calculated according to the following equation:

$E'_f = E[(Pt_5/Fe_{3-x}O)] + E[Fe\,(g)] - E[(Pt_5/Fe_3O_4)]$ \hspace{1cm} (7)

**Table S11.** Fe vacancy energy ($E_f$, eV) and total spin magnetic moment per unit cell (M, $\mu_B$).

|  | Figure | $E'_f$ | $M_{total}$ |
|---|---|---|---|
| $V_{Fe1}$ | S15(a) | 6.70 | 4.84 |
| $V_{Fe2}$ | S15(b) | 6.90 | 3.64 |
| $V_{Fe3}$ | S15(c) | 7.69 | 2.52 |
| $V_{Fe4}$ | S15(d) | 6.37 | 2.87 |

We then consider the preferred adsorption sites for a single Fe atom on the various isomers of $Pt_5$ in the gas phase. Altogether, we have considered 20 possible isomers. Table S12 reports the optimized structures and their relative energy ($E_R$). The binding energy of Fe to $Pt_5$ in the most stable isomer is -0.99 eV.

**Table S12.** Relative energy ($E_R$) of an adsorbed Fe atom on $Pt_5$ clusters.

| Structure | | $E_R$, eV | Structure | | $E_R$, eV |
|---|---|---|---|---|---|
| 1 | 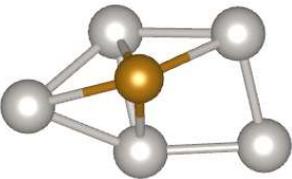 | 0.00 | 11 | 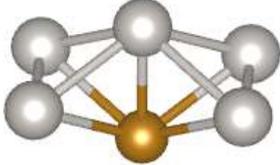 | 0.82 |
| 2 | 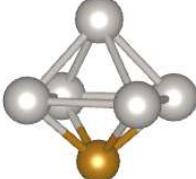 | 0.06 | 12 | 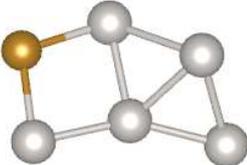 | 0.95 |
| 3 | 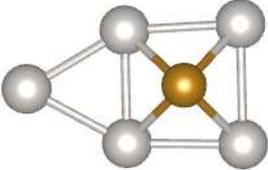 | 0.16 | 13 | 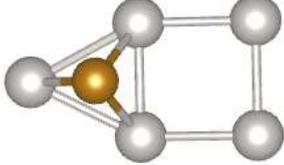 | 1.04 |
| 4 | 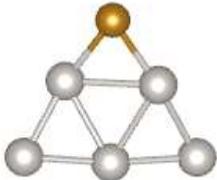 | 0.23 | 14 | 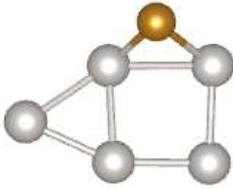 | 1.10 |
| 5 | 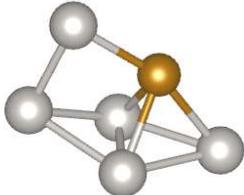 | 0.28 | 15 | 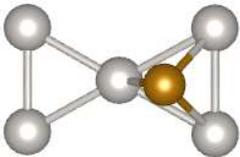 | 1.49 |
| 6 | 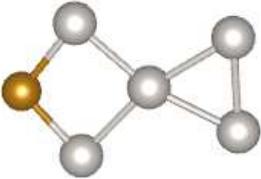 | 0.39 | 16 | 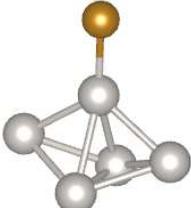 | 1.76 |
| 7 | 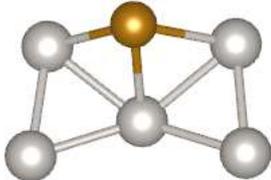 | 0.49 | 17 | 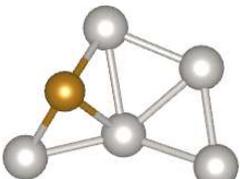 | 2.39 |
| 8 | 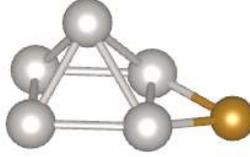 | 0.56 | 18 | 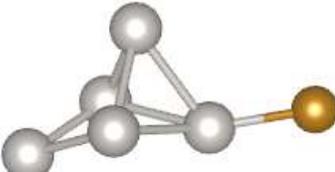 | 2.71 |
| 9 | 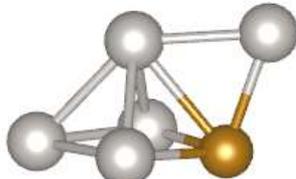 | 0.69 | 19 | 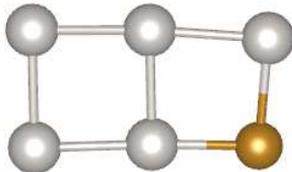 | 2.74 |
| 10 | 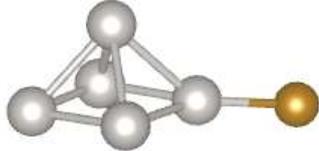 | 0.73 | 20 | 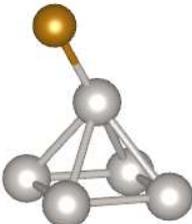 | 2.76 |

We now consider the energetic cost of displacing the Fe1 (tetrahedral) and Fe4 (octahedral) atoms, indicated in Figure S15, from the support and adsorbing it on the $Pt_5$ cluster. The Fe atom has been re-adsorbed on various sites of the supported $Pt_5$ cluster, starting from the most stable structure (Figure 4 (a) in the main text). Figure S17 (a)-(c) and (e)-(f) shows the optimized structures. Based on our results for the $Pt_5Fe$ gas phase isomers (Table S12), the re-adsorption of the Fe atom on the triangular facet of a supported $Pt_5$(i) cluster, Figure 4 (c), has also been studied. Figure S17 (d) shows the optimized structure.

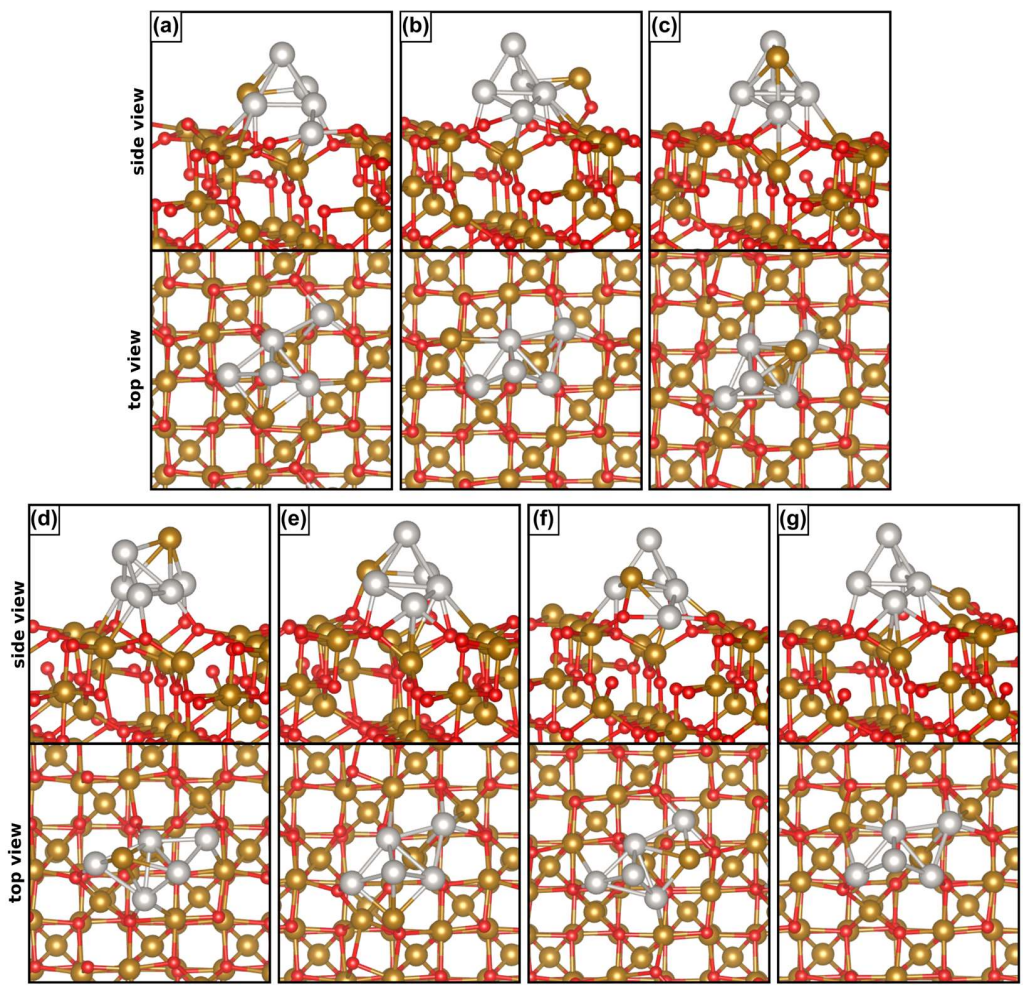

**Figure S17.** Various isomers of $Pt_5Fe$ supported on $Fe_3O_4$.

**Table S13.** Fe spillover energy ($E_{SO}$, eV) and its site shown in Figure S15 and total spin magnetic moment per unit cell (M, $\mu_B$).

| Figure | Fe site | $E_{SO}$[a] | $M_{total}$ |
|---|---|---|---|
| S16 (a) | Fe1 | 2.38 | 4.83 |
| S16 (b) | Fe1 | 2.53 | 4.07 |
| S16 (c) | Fe4 | 2.61 | 3.23 |
| S16 (d) | Fe4 | 3.18 | 1.75 |
| S16 (e) | Fe4 | 3.38 | 4.17 |
| S16 (f) | Fe1 | 4.17 | 3.57 |
| S16 (g)[b] | Fe4 | 0.14 | 3.24 |

[a] $Pt_5/Fe_3O_4 \rightarrow Fe\text{-}Pt_5/Fe_{3-x}O_4$ 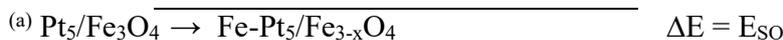 $\Delta E = E_{SO}$

[b] In this case, during the optimization, the Fe atom adsorbed on a triangular facet of $Pt_5$ moved back to the $Fe_3O_4$ surface.

In all cases, the Fe spillover implies a remarkable cost (between 2.38 eV and 4.17 eV, depending on the initial and final sites). Case (g) looks like an outlier displaying an almost thermoneutral formation energy, but is in fact a failed attempt, where the Fe ion spontaneously left the $Pt_5$ cluster and moved back to its lattice site during the relaxation.

Next, we consider the energetic cost of displacing the Fe1 (tetrahedral) or Fe4 (octahedral) atoms (see Figure S15) from the support and adsorbing them on the $Pt_5O_3$ cluster. The Fe atom has been re-adsorbed on the triangular facet of the supported $Pt_5O_3$ cluster starting from the most stable structure shown in Figure S13 (a). The results are summarized in Figure S18 and Table S14, indicating a non-favorable process.

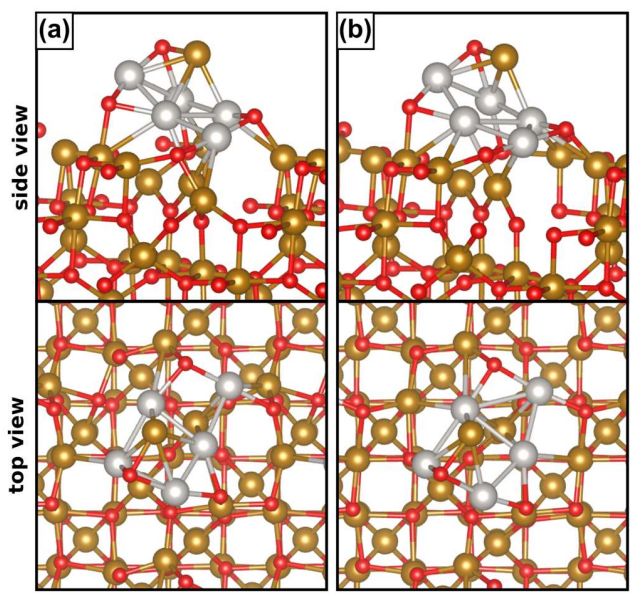

**Figure S18.** Various isomers of Pt$_5$O$_3$Fe supported on Fe$_3$O$_4$.

**Table S14.** Fe spillover energy ($E_{SO}$, eV) and its site shown in Figure S15 and total spin magnetic moment per unit cell (M, $\mu_B$).

| Figure | Fe site | $E_{SO}^{(a)}$ | $M_{total}$ |
| --- | --- | --- | --- |
| S17 (a) | Fe4 | 2.18 | 2.59 |
| S17 (b) | Fe1 | 3.70 | 4.12 |

(a) $E_{SO}$ refers to the process: Pt$_5$/Fe$_3$O$_4$ → Pt$_5$O$_3$Fe/Fe$_{3-x}$O$_{4-y}$